\documentclass[12pt,draftclsnofoot, onecolumn]{IEEEtran}
\usepackage[T1]{fontenc}
\usepackage{amsthm,amsmath,amssymb,mathtools,bm}
\usepackage{stfloats}
\usepackage{graphicx}
\usepackage{subfigure}
\usepackage{cite}	
\theoremstyle{remark}

\newtheoremstyle{MyThmStyle}
{}
{}
{}
{}
{\em}
{:}
{ }
{\hspace{1em}\thmname{#1}\thmnumber{ #2}\thmnote{ (#3)}}

\theoremstyle{MyThmStyle}
\newtheorem{myremark}{Remark}
\newtheorem{mytheorem}{Theorem}
\newtheorem{myproposition}{Proposition}
\newtheorem*{myproof}{Proof}

\ifCLASSINFOpdf
\else
\fi

\begin{document}
	
	\title{On the Characterizations of OTFS Modulation over
		Multipath Rapid Fading Channel}
	
	\author{Haoyan~Liu,
		Yanming~Liu,
		Min~Yang,
		and Qiongjie~Zhang
		\thanks{The authors are the School of Aerospace Science and Technology, Xidian University, Xi’an 710071, China.}
	}
	
	\maketitle
	
	\begin{abstract}
		
		Orthogonal time frequency space (OTFS) modulation has been confirmed to provide significant performance advantages against Doppler in high-mobility scenarios. The core feature of OTFS is that the time-variant channel is converted into a non-fading 2D channel in the delay-Doppler (DD) domain so that all symbols experience the same channel gain. In now available literature, the channel is assumed to be quasi-static over an OTFS frame. As for more practical channels, the input-output relation will be time-variant as the environment or medium changes. In this paper, we analyze the characterizations of OTFS modulation over a more general multipath channel, where the signal of each path has experienced a unique rapid fading. First, we derive the explicit input-output relationship of OTFS in the DD domain for the case of ideal pulse and rectangular pulse. It is shown that the rapid fading will produce extra Doppler dispersion without impacting on delay domain. We next demonstrate that OTFS can be interpreted as an efficient time diversity technology that combines space-time encoding and interleaving. Simulation results reveal that OTFS is insensitive to rapid fading and still outperforms orthogonal frequency-division multiplexing (OFDM) in these types of channels.

	\end{abstract}
	

	\IEEEpeerreviewmaketitle
	
	\section{Introduction}
	\IEEEPARstart{I}{n} mobile wireless communication, linear time-variant (LTV) channels are typically represented in terms of time and frequency shifts on the transmitted signal due to the presence of multipath propagation and motion or carrier frequency offsets. For multiplexing transmission in such channels, a challenging task is to design an optimal modulation scheme to avoid mutual crosstalk between different time slots and subcarriers\cite{1347350}. It is well-known that orthogonal frequency-division multiplexing (OFDM) is robust to inter-symbol interference (ISI) for data transmission over multipath fading channels. However, it will experience significant performance degradation in high-mobility scenarios. The frequency dispersion induced by Doppler shift destroys the orthogonality condition and yields severe inter-carrier interference (ICI).
	
	Recently, a new modulation technique called orthogonal time frequency space (OTFS) has been proposed \cite{7925924,1802.02623,8058662}. The most remarkable difference from conventional time-frequency (TF) modulation is that OTFS equivalently modulates information symbols in the delay-Doppler (DD) domain. As a result, the aforementioned LTV channels are converted into the time-invariant channels in the DD domain, and all symbols over a transmission frame experience the same channel gain. It has been shown that OTFS greatly outperforms OFDM for user's velocity up to 500 km/h in LTE system \cite{7925924}. 
	
	Because of the enormous advantage against Doppler spread, OTFS has attracted widespread attention as a next generation modulation scheme. To analyze the gain obtained by OTFS over OFDM, OTFS was described as block-OFDM with a cyclic prefix and time interleaving \cite{2001.02446}. In \cite{1802.02623}, OTFS was considered to exploit channel diversity gain, and formal analysis of the diversity order was provided in \cite{8756831,8686339}. In \cite{8918014,8859227}, two types of low-complexity minimum mean square error (MMSE) detectors were proposed, respectively. As for maximum likelihood (ML) detection, \cite{9082873} proposed a variational Bayes (VB) approach as an approximation of the optimal ML detection. Based on the sparsity of channel state information (CSI) and Gaussian approximation of the interference terms, a low-complexity message passing (MP) detector was proposed in \cite{8424569}, while the detailed input-output relation of OTFS in DD domain was also formulated. Moreover, OTFS has also proven to be as applicable to many communication systems as OFDM. \cite{8761362} investigated the 3D structured sparse channel of multiple-input multiple-output OTFS (MIMO-OTFS) and proposed a channel estimation method. In\cite{9109735}, it is showed that OTFS could enable to efficiently achieve full information rate of the modulation and near-optimal radar estimation performance.
	
	The above literature primarily focus on quasi-static channels, put it another way, the Doppler of each path is a constant over an OTFS frame. However, the rapid motion will cause Doppler scaling on the transmitted signal with time. In addition, the doubly dispersive fading model is not well matched to all possible wireless channels. When the interaction between signal and channel is specific to the environment or medium that the signal propagates in, such as electromagnetic (EM) waves in the ionosphere, the time-scale will change and cannot be directly approximated by frequency shifts \cite{995064}. In this case, a better matched representation in terms of time scale and frequency scale change should be expected.
	
	In this paper, we consider a more generalized multipath channel model with rapid fading and study the characterizations of OTFS over this channel. For the purpose of comparison with the aforementioned LTV model, we assume that the signal propagating in each path has experienced a unique time-variant fading in addition to delay and Doppler spread, where the time-variant fading is used to describe scale change. The core contributions in this work can be summarized as follows.
	
	\begin{itemize}
		\item We first derive the input-output relation in a single-input single-output (SISO) system over multipath rapid fading channel. Weyl-Heisenberg (WH) Basis is a universal scheme in transmission systems, and it has been extensively studied over the doubly dispersive fading channel \cite{1347350,730463,5089511,STROHMER2001243}. Since the rapid fading will further destroy the orthogonality of WH basis, we commence from the interaction between WH basis and multipath rapid fading channel, and formulate an explicit explanation. Then, we demonstrate the corresponding input-output relation in the case of ideal pulse and rectangular pulse in DD domain, respectively. It is observed that the rapid fading has no effect on the sparsity of the delay domain, while it will further disperse the Doppler domain in a convolutional manner.
		\item We prove that OTFS as an encoding and interleaving technique contains inherent time diversity order, where the Discrete Fourier Transform (DFT) matrix and multipath delays perform the encoding, and the equivalently transpose operation corresponding to block-OFDM perform the interleaving. In \cite{661517}, a general framework for analyzing the performance of space-time coding was proposed. We follow this framework and deduce that the asymptotic time diversity order of OTFS is $PN$, where $P$, $N$ represent the number of Doppler lattices and paths. Therefore, the bit-error-rate (BER) performance of OTFS asymptotically tends to Gaussian channel with increasing $P$ and $N$. Furthermore, The simulation results show that OTFS still outperforms OFDM in rapid fading channels.
		
	\end{itemize}

	The rest of the paper is organized as follows. In Section II, we briefly review the WH basis and OTFS modulation. In Section III, we derive the input-output relation over multipath rapid fading channels. In Section IV, the time diversity order of OTFS is analyzed. Simulation results are given in Section V. Conclusions are finally presented in Section VI.
	
	\section{System Description}
	\subsection{Weyl-Heisenberg Basis}
	The input-output relation of a doubly dispersive channel can be modeled as a linear transformation operator $\bm{{\cal H}}$ and given by
	\begin{equation}
	\begin{aligned}
	r(t) &= (\bm{{\cal H}}s)(t) = \int_{\tau} \widetilde{h} (t,\tau)x(t-\tau) d\tau \\ &= \int_{\tau} \int_{\nu} h(\tau,\nu) s(t-\tau) e^{j2\pi\nu (t-\tau)} d\tau d\nu,
	\label{eq1}
	\end{aligned}
	\end{equation}
	namely, the output signal $r(t)$ can be described as a weighted superposition of time and frequency shifted copies of the input signal $s(t)$ in terms of the channel delay-Doppler spreading function $h(\tau,\nu)$. For the convenience of analysing, the additive noise is ignored. In practice, a wide variety of physical environments satisfies the wide-sense stationary uncorrelated scattering (WSSUS) assumption, i.e., $h(\tau, \nu)$ is uncorrelated in different delays and Doppler shifts:
	\begin{equation}
	\mathbb{E}[h(\tau,\nu)h^{*}(\tau^{\prime},\nu^{\prime})] = S(\tau,\nu)\delta(\tau - \tau^{\prime})\delta(\nu-\nu^{\prime})
	\end{equation}
	where $S(\tau,\nu)$ denotes the scattering function of the channel considered to be normalized
	\begin{equation}
	\int_{\tau} \int_{\nu} S(\tau,\nu) d\tau d\nu = 1.
	\end{equation}
	
	In the digital communications, the eigendecomposition of the operator $\bm{{\cal H}}$ is significant for modulation design \cite{MATZ20111,4489232}. The $\lambda_{k}$ and $u_{k}(t)$ are the channel eigenvalues and eigenfunctions defined by
	\begin{equation}
	(\bm{{\cal H}}u_{k})(t)=\int_{t^{\prime}} \widetilde{h}(t,t^{\prime})u_{k}(t^{\prime}) dt^{\prime}=\lambda_{k} u_{k}(t).
	\end{equation}
	where the set $ \{ u_{k}(t) \}^{\infty}_{k=0}$ constitutes a complete orthonormal basis in $L^{2}(\mathbb{R})$ space. Therefore, the transmitted symbols $X_{k}$ can be modulated onto the orthonormal basis, which yields the transmitted signal $s(t) = \sum_{k} X_{k} u_{k}(t)$. Then at the receiver, demodulation can be perfectly performed through projecting $r(t)$ onto the basis (or called matched filter processing)
	\begin{equation}
	\begin{aligned}
	Y_{k} &= \langle (\bm{{\cal H}}s)(t),u_{k}(t) \rangle \\ 
	&=\sum_{k^{\prime}} X_{k^{\prime}} \langle (\bm{{\cal H}}u_{k^{\prime}})(t),u_{k}(t) \rangle =\lambda_{k} X_{k}.
	\label{eq5}
	\end{aligned}
	\end{equation}
	
	The above channel diagonalization will enable a one-tap detector and equalizer in accordance with the pretty simple input-output relation, so (\ref{eq5}) produces an ideal communication system. However, it cannot be applied to practical systems. This requires the receiver and transmitter to dynamically accommodate the transmitted basis to the channel eigenfunctions based on the full knowledge of the channel realization. Instead, for the purpose of coping with various channels, WH basis is often regarded as approximate eigenfunction for transmission used in most linear modulation schemes. The WH basis is defined as
	
	\begin{equation}
	g_{nm}(t) = g(t-nT)e^{j2\pi m \Delta f\left(t-nT\right)}, \quad (n,m) \in \mathbb{Z}^{2}
	\end{equation}
	where $g(t) \in L^{2}(\mathbb{R})$ is the normalized prototype pulse, $T$ is the symbol period and $\Delta f$ is the carrier separation. The $g_{nm}(t)$ can be construed as a time-frequency shifted version of $g(t)$ so that the WH basis composed of the triple $\{g(t),T,\Delta f\}$ has an intuitional physical interpretation for pratical system. The product $T\Delta f$ determines the completeness of the WH basis. In this paper, we will primarily focus on $T\Delta f=1$ and it yields complete orthonormal transmission scheme well known as OFDM. Unfortunately, the complete WH basis have necessarily bad time-frequency localization.
	
	As approximate eigenfunction of the operator $\bm{{\cal H}}$, the ideal diagonalization cannot be performed while the delay and Doppler exist in the channel, i.e., the orthogonality of WH basis will be destroied. Hence, it will result in a more complicated input-output relation than (\ref{eq5}) because of the presence of the off-diagonal terms. After projecting $r(t)$ onto the WH basis, the received symbol can be rewritten as
	\begin{equation}
	Y_{nm} = H_{nm,nm}X_{nm} + \sum_{n^{\prime} \neq n \, \textrm{or} \, m^{\prime} \neq m} H_{nm,n^{\prime}m^{\prime}}X_{n^{\prime}m^{\prime}}.
	\label{eq7}
	\end{equation}
	In (\ref{eq7}), the second term represents the interfence caused by the off-diagonal terms towards symbol $X_{nm}$ and the $H_{nm,n^{\prime}m^{\prime}}$ is given by
	\begin{equation}
	\begin{aligned}
	&H_{nm,n^{\prime}m^{\prime}} = \langle (\bm{{\cal H}}g_{n^{\prime}m^{\prime}})(t),g_{nm}(t) \rangle \\ &= \int_{\tau} \int_{\nu} h(\tau,\nu) A_{g}\left((n-n^{\prime})T-\tau,(m-m^{\prime})\Delta f - \nu\right) e^{j2\pi (m^{\prime}f+\nu)\left((n-n^{\prime})T - \tau\right)}e^{j2\pi \nu n^{\prime}T} d\tau d\nu,
	\end{aligned}
	\end{equation}
	with the ambiguity function
	\begin{equation}
	\begin{aligned}
	A_{g}(\tau,\nu) &= \langle g(t),g(t-\tau)e^{j2\pi\nu(t-\tau)} \rangle \int_{t} g(t)g^{*}(t-\tau)e^{-j2\pi\nu(t-\tau)}dt.
	\end{aligned}
	\end{equation}
	The $A_{g}(\tau,\nu)$ formulates the correlation of the signal with a time and frequency shifted version of itself and satisfies the following properties
	\begin{equation}
	A_{g}(\tau,\nu) \leq A_{g}(0,0) = \|g\|^{2} = 1
	\end{equation}
	and
	\begin{equation}
	A_{g}(nT,m\Delta f) = 0, \quad \forall (n,m) \neq (0,0).
	\end{equation}
	For the first term of (\ref{eq7}), each symbol $X_{nm}$ suffers a unique $H_{nm,nm}$, so it is vulnerable to additive noise in deep fading case. In the second term, two kinds of interference are contained. The interference rendered by Doppler at different frequencies $m^{\prime} \neq m$ but the same time slot $n$ is named as ICI. For another rendered by delay between adjacent time slots $n^{\prime} \neq n$ is called ISI. The better robustness against channel dispersion could be carried out through designing well localized prototype pulse $g(t)$. In OFDM, the ISI is canceled via adding appropriate CP; however, the ICI cannot be avoided and arouse high BER.
	
	\subsection{OTFS Modulation}
	
	The significant feature of OTFS is to modulate data symbols (e.g., QAM symbols) in delay-Doppler domain. Specifically, the data sequence is first rearranged into a $N \times M$ lattice, where $N$ and $M$ are the numbers of points of the lattice along the delay and Doppler axis. The symbols $x_{kl}$ residing in delay-Doppler domain are converted into time-frequency domain through the 2D inverse symplectic finite Fourier transform (ISFFT)
	\begin{equation}
	X_{nm} = \frac{1}{\sqrt{NM}}\sum^{N-1}_{n=0}\sum^{M-1}_{m=0} x_{kl}e^{j2\pi \left(\frac{nk}{N}-\frac{ml}{M}\right)}.
	\end{equation}
	This procedure can be treated as a pre-processing compared with the traditional OFDM. For multiplexing at air interface, the symbols $X_{nm}$ are further modulated on the WH basis
	\begin{equation}
	s(t) = \sum_{n=0}^{N-1} \sum_{m=0}^{M-1} X_{nm}g(t-nT)e^{j2\pi m\Delta f(t-nT)}.
	\end{equation}
	Assume that there are $P$ paths in the channel, where each path is associated with a delay $\tau_{i}$, Doppler $\nu_{i}$ and a fade coefficient $h_{i}$. In addition, the channel is supposed to be underspread, i.e., $\tau_{\textrm{max}} < T$ and $\nu_{\textrm{max}} < \Delta f$. The representation of the spreading function $h(\tau,\nu)$ is given as
	\begin{equation}
	h(\tau, \nu)=\sum_{i=1}^{P} h_{i} \delta\left(\tau-\tau_{i}\right) \delta\left(\nu-\nu_{i}\right).
	\end{equation}
	Hence, the sum operator is substituted for the integral operator in (\ref{eq1}) and the received signal with additive Gaussian noise $n(t)$ can be written as
	\begin{equation}
	\begin{aligned}
	r(t) = (\bm{{\cal H}}s)(t) + n(t) = \sum_{i=1}^{P} h_{i}s(t-\tau_{i})e^{j2\pi \nu (t-\tau_{i})} + n(t).
	\end{aligned}
	\end{equation}	
	After implementing match filter, the received $Y_{nm}$ is consistent with (\ref{eq7}). If the prototype pulse $g(t)$ is specified to be ideal localized, it yields $A_{g}\left(nT\pm\tau_{max},m\Delta f\pm\nu_{max}\right)=\delta[n]\delta[m]$, and the received symbols $Y_{nm}$ will equivalently simplify as (\ref{eq5}). Then, the symplectic finite Fourier transform (SFFT) is applied to obtain the demodulated data as
	\begin{equation}
	y_{kl}=\frac{1}{\sqrt{N M}} \sum_{n=0}^{N-1} \sum_{m=0}^{M-1} Y_{nm} e^{-j 2 \pi\left(\frac{n k}{N}-\frac{m l}{M}\right)} + w_{kl},
	\label{eq16}
	\end{equation}
	where the $w_{kl}$ is the noise in delay-Doppler domain. Notably all the transforms are orthonormal in the system, so $w$ follow the same Gaussian distribution as $n(t)$. 
	
	From the above transforms, the resolutions in delay and Doppler axes are $1/M\Delta f$ and $1/NT$ respectively, thus the dealy $\tau_{i}$ and Doppler $\nu_{i}$ can be rewritten in index form, $\tau_{i}=\frac{l_{\tau_{i}}}{M \Delta f}$ and  $\nu_{i}=\frac{k_{\nu_{i}}+\kappa_{\nu_{i}}}{N T}$. The $l_{\tau_{i}}$ and $k_{\nu_{i}}$ represent the integral indexes of delay and Doppler tap, and the $\kappa_{\nu_{i}}$ represents the fractional part. Taking the ideal prototype pulse, the end-to-end system can be formulated as
	\begin{equation}
	y_{kl} = \frac{1}{N}\sum_{k^{\prime}=0}^{N-1} \sum_{l^{\prime}=0}^{M-1} x_{k^{\prime}l^{\prime}}h_{\omega}[k-k^{\prime},l-l^{\prime}],
	\label{eq17}
	\end{equation}
	and the $h_{\omega}$ is given by
	\begin{equation}
	\begin{aligned}
	h_{\omega}[k-k^{\prime},l-l^{\prime}] =M\sum_{i=1}^{P}h_{i} e^{-j2\pi\frac{l_{\tau_{i}}(k_{\nu_{i}}+\kappa_{\nu_{i}})}{NM}}\beta_{i}(k-k^{\prime})  \delta\left([l-l^{\prime}-l_{\tau_{i}}]_{M}\right),
	\label{eq18}
	\end{aligned}
	\end{equation}
	where
	\begin{equation}
	\beta_{i}(k-k^{\prime}) =  \sum_{n=0}^{N-1} e^{-j \frac{2\pi n}{N}(k-k^{\prime}-k_{\nu_{i}}-\kappa_{\nu_{i}})}.
	\end{equation}
	and $[\cdot]_{M}$ represents mod $M$ operation. Clearly, (\ref{eq17}) describes a 2D convolution profile, which reveals that each transmitted symbol suffers all the channel response. As a consequence, OTFS modulation naturally takes advantage of all the diversity paths in the channel through implementing maximum likelihood detector, which makes OTFS outperform OFDM. Moreover, it is conveniently to cope with the channel estimation on account of the sparsity of the $h_{\omega}$.

	\section{Input-Output Relation over Multipath Rapid Fading Channel}
	In this section, we consider a more general LTV channel \cite{850678}: when the transmitted signal suffers a rapid fading $\gamma^{i}(t)$ in each propagation paths, where both the amplitude and phase are functions of $t$. Without loss of generality, the received signal $r(t)$ is modeled by the mathematical equation
	\begin{equation}
	r(t) = (\bm{{\cal H}} \bm{{\cal D}}s)(t) \triangleq \sum_{i=1}^{p} {\gamma}^{i}(t)s(t-\tau_{i})e^{j2\pi \nu(t-\tau_{i})}.
	\end{equation}
	Undoubtedly, the operator $\bm{{\cal D}}$ will further destroy the orthogonality of WH basis and impact on the input-output relation in delay-Doppler domain.
	
	\begin{myremark}
		There have several different representations of the input-output relation, which differ by the propagation of EM waves in various scenarios. If each path has its specific propagation characteristics, e.g. underwater environments, the received signal corresponds to a linear combination of $P$ independent distortions of the transmitted signal. In another case, for time-variant propagation medium just existing around the transmitted or received antenna, the multipath signal is considered to experience the same dispersion and $\gamma(t)$ can be make no distinction. Such examples can be found in plasma sheath channel \cite{4103818}. Furthermore, the response of a system to a unit pulse at time $t$ or $t-\tau$ determines whether there exists a translation in $\gamma(t)$. If the signal undergoes $\gamma(t)$ and then propagates in multipath, the translation needs to be taken into account, or vice versa. The final results in different cases can be analogously derived in conformity to the corresponding definition.
	\end{myremark}
	
	First, we investigate the time-frequency CIR to understand the interaction between the WH basis and the LTV channel
	\begin{equation}
	\begin{aligned}
	&C_{nm,n^{\prime}m^{\prime}} = \langle (\bm{{\cal H}} \bm{{\cal D}}g_{n^{\prime}m^{\prime}})(t), g_{nm}(t) \rangle \\
	& \quad = \sum_{i=1}^{P} \int_{t} \gamma^{i}(t) g_{n^{\prime}m^{\prime}}(t-\tau_{i})e^{j2\pi \nu_{i}(t-\tau_{i})} g^{*}_{nm}(t) dt.
	\label{eq20}
	\end{aligned}
	\end{equation}
	Here, $C_{nm,n^{\prime}m^{\prime}}$ could be problematic to directly derive on account of the integral operator even if the explicit profiles of $g(t)$ and $\gamma(t)$ are given. To deal with this problem, we use a discretization method based on the properties of WH basis instead.
	
	From the completeness of the WH basis, a signal $f(t) \in L^{2}(\mathbb{R})$ can be reconstructed from its expansion coefficients $\{ \langle f(t), g_{nm}(t) \rangle \}$, which is accomplished according to \cite{frame}
	\begin{equation}
	f(t) =  \sum_{n,m} \langle f(t), g_{nm}(t) \rangle \ g_{nm}(t).
	\end{equation}
	In this way, the $g_{nm}(t)$ accompanied by delay $\tau_{i}$ and Doppler $\nu_{i}$ can be rewritten as
	\begin{equation}
	g_{nm}(t-\tau)e^{j2\pi \nu(t-\tau)} = \sum_{n^{\prime},m^{\prime}} H^{i}_{nm,n^{\prime}m^{\prime}} g_{n^{\prime}m^{\prime}}(t),
	\label{eq22}
	\end{equation}
	where
	\begin{equation}
	\begin{aligned}
	H^{i}_{nm,n^{\prime}m^{\prime}} =  A_{g}\left((n-n^{\prime})T-\tau_{i},(m-m^{\prime})\Delta f - \nu_{i}\right)  e^{j2\pi \left(m^{\prime}\Delta f+\nu_{i}\right)\left((n-n^{\prime})T - \tau_{i}\right)}e^{j2\pi \nu n^{\prime}T}.
	\label{eq23}
	\end{aligned}
	\end{equation}
	Substituting the reconstructed form into (\ref{eq20}), the $C_{nm,n^{\prime}m^{\prime}}$ is calculated by
	\begin{equation}
	\begin{aligned}
	C_{nm,n^{\prime}m^{\prime}} = \langle \sum_{i=1}^{p} \gamma^{i}(t) \sum_{n^{\prime\prime},m^{\prime\prime}} H^{i}_{n^{\prime}m^{\prime},n^{\prime\prime}m^{\prime\prime}}g_{n^{\prime\prime}m^{\prime\prime}}(t), g_{nm}(t)\rangle \\
	= \sum_{i=1}^{p} \sum_{n^{\prime\prime},m^{\prime\prime}} H^{i}_{n^{\prime}m^{\prime},n^{\prime\prime}m^{\prime\prime}}\langle \gamma^{i}(t)g_{n^{\prime\prime}m^{\prime\prime}}(t), g_{nm}(t)\rangle.
	\label{eq24}
	\end{aligned}
	\end{equation}
	(\ref{eq24}) states that the $C_{nm,n^{\prime}m^{\prime}}$ can be alternatively formulated as a linear accumulation of a set of inner products and the corresponding $H^{i}_{nm,n^{\prime}m^{\prime}}$. Hence, the effects of delay and Doppler are removed out of the integral and produce a tractable inner product term. The inner products are the CIR of $\gamma^{i}(t)$, which describe the dispersion generated by the $\gamma^{i}(t)$ on each point of time-frequency lattice.
	
	\begin{mytheorem}
		In WH system, the channel impulse
		response (CIR) of time-variant fading $\gamma(t)$ among time-frequency lattice can be characterised as
		\begin{equation}
		\begin{aligned}
		\langle \gamma(t)g_{n^{\prime}m^{\prime}}(t), g_{nm}(t)\rangle
		= 
		\begin{dcases}
		\frac{1}{M}\sum\limits_{u=0}^{M-1} \overline{\gamma}_{n}(u) e^{-j2\pi\frac{u}{M}(m-m^{\prime})} \quad &n=n^{\prime},\\ 
		0 \quad &\mathrm{otherwise}.
		\end{dcases} 
		\end{aligned}
		\label{eq25}
		\end{equation}
		where the $\overline{\gamma}_{n}(u)$ is the discretization of $\gamma(t)$ with sampling interval $1/M\Delta f$ over $n$-th symbol period.
		\label{th1}
	\end{mytheorem}
	\begin{myproof}
		See Appendix A.
	\end{myproof}
	Theorem \ref{th1} provides an explicit explanation on the interaction between WH basis and $\gamma(t)$. It turns out that $\gamma(t)$ only leads to dispersion in frequency domain, and the corresponding value is consistent with $M$-points DFT of $[\overline{\gamma}_{n}(0),\overline{\gamma}_{n}(1),\cdots,\overline{\gamma}_{n}(M-1)]$ being divided by $M$. For $m=m^{\prime}$, the CIR corresponds to the zero-frequency component. In addition, as the $m^{\prime}$ moves, the CIR will be circular shifted along frequency axis. 
	
	With respect to fixed $n$ and $m$, the frequency response is the circular convolution of the original CIR caused by Doppler with the discrete spectrum of $\gamma(t)$. Since the different prototype pulse will produce specific $H^{i}$ and input-output relation in delay-Doppler domain, in the following subsections, we will discuss the cases for ideal and rectangular pulse respectively.
	
	\subsection{Ideal Pulse}
	In (\ref{eq23}), it indicates that $H^{i}_{nm,n^{\prime}m^{\prime}}$ is non-zero only at $n^{\prime}=n, m^{\prime}=m$ for ideal pulse. Hence, combined with Theorem \ref{th1}, the following relation holds
	\begin{equation}
	Y_{nm} = \sum_{m^{\prime}=0}^{M-1} C_{nm,nm^{\prime}}X_{nm^{\prime}} + W_{nm},
	\label{eq26}
	\end{equation}
	where
	\begin{equation}
	C_{nm,nm^{\prime}} = \frac{1}{M} \sum_{i=1}^{p} H^{i}_{nm^{\prime},nm^{\prime}} \sum\limits_{u=0}^{M-1} \overline{\gamma}_{n}^{i}(u) e^{-j2\pi\frac{u}{M}(m-m^{\prime})}.
	\label{eq27}
	\end{equation}
	Then, by implementing SFFT, the final delay-Doppler CIR and the end-to-end system model can be obtained.
	\begin{myproposition}
		The delay-Doppler CIR with ideal pulse can be characterized as
		\begin{equation}
		\begin{aligned}
		h^{\mathrm{DD}}_{kl,k^{\prime}l^{\prime}} = M \sum_{i=1}^{P} e^{-j2\pi\frac{l_{\tau_{i}}(k_{\nu_{i}}+\kappa_{\nu_{i}})}{NM}} \alpha_{i}^{\mathrm{ideal}}(k,k^{\prime},l) \delta\left([l-l^{\prime}-l_{\tau_{i}}]_{M}\right),
		\end{aligned}
		\end{equation}
		and the demodulated signal $y_{kl}$
		\begin{equation}
		\begin{aligned}
		y_{kl} =\frac{1}{N} \sum_{i=1}^{P}  e^{-j2\pi\frac{l_{\tau_{i}}(k_{\nu_{i}}+\kappa_{\nu_{i}})}{NM}} \sum_{k^{\prime}=0}^{M-1} \alpha_{i}^{\mathrm{ideal}}(k,k^{\prime},l)  x\left[k^{\prime},[l-l_{\tau_{i}}]_{M}\right] + w_{kl},
		\label{eq29}
		\end{aligned}
		\end{equation}
		where
		\begin{equation}
		\alpha_{i}^{\mathrm{ideal}}(k,k^{\prime},l)=\underbrace{\frac{1}{N}\sum_{n=0}^{N-1} \overline{\gamma}_{n}^{i}(l) e^{-j2\pi k\frac{n}{N}}}_{\text{extra Doppler interference}} \circledast\beta_{i}(k-k^{\prime}),
		\end{equation}
		and $\circledast$ denotes circular convolution.
		\label{th2}
	\end{myproposition}
	\begin{myproof}
		See Appendix B.
	\end{myproof}
	We observe that the sparsity of delay domain can be still guaranteed as the $\gamma(t)$ simply impacts on Doppler domain. Concerning to each path, the $h^{\mathrm{DD}}$ in Doppler axis can be decomposed into two parts and formulated as a column-wise circular convolution of the original Doppler response with the extra Doppler dispersion. Similar with (\ref{eq25}), the extra Doppler dispersion is equivalent to $N$-points DFT of $\boldsymbol{\gamma}_{i,l}=[\overline{\gamma}^{i}_{0}(l),\overline{\gamma}^{i}_{1}(l),\cdots,\overline{\gamma}^{i}_{N-1}(l)]^{^{\rm{T}}}$. That is, each original response will continue to disperse towards both sides of the Doppler axis in the form of discrete spectrum of $\boldsymbol{\gamma}_{i,l}$. Note that the sampling interval here is $1/T$ (or $\Delta f$) in time, hence the quantization step is $1/NT$ (or $\Delta f / N$) in frequency. As a result, the subcarrier separation dominates the highest cut-off frequency and the length of OTFS frame dominates the frequency resolution. When fractional Doppler is zero, $h^{\mathrm{DD}}$ is consistent with the discrete spectrum, where the zero-frequency component appears at point $k^{\prime}=k-k_{\nu_{i}}$. Therefore, the parameters in OTFS are closely related to the Doppler response as well. This will be discussed latter.
	
	Being different from $h_{\omega}$, which is same for all transmitted symbols, $\gamma^{i}(t)$ yields particular interferences for each symbol on delay axis if without any restrictions on $\gamma^{i}(t)$. At the receiver, this results that the amount of coefficients being estimated is $N-1$ times more than before. If the $\boldsymbol{\gamma}_{i,l}$ is taken to be a wide sense stationary and ergodic distortion process, $h^{\mathrm{DD}}$ has the same amplitude for each $x_{kl}$. Only when $\gamma^{i}(t)$ satisfies blocked variation, i.e., the $\gamma^{i}(t)$ approximately remains fixed on a symbol period, all transmitted symbols will suffer the identical $h^{\mathrm{DD}}$.
	
	\subsection{Rectangular Pulse}
	
	Unlike the ideal pulse is adopted to analyze the bound on performance, the rectangular pulse is commonly used in practical system. With the assumption that $\tau_{max} < T$, $Y_{nm}$ will only encompass ISI from the previous symbol period $n-1$. It yields that $C_{nm,n^{\prime} m^{\prime}}$ is non-zero when $n^{\prime}=n$ and $n^{\prime}=n-1$. Resemble (\ref{eq27}), $C_{nm,n^{\prime} m^{\prime}}$ can be deduced by using (\ref{eq24}) for an explicable explanation. Nevertheless, considering that rectangular pulse has a tractable form, we adopted a more straight approach (by using (\ref{eq20})) to avoid introducing additional sum operators. Since the received signal is sampled at intervals of $1/M\Delta f$, ICI alternatively becomes
	\begin{equation}
	\begin{aligned}
	C_{nm,nm^{\prime}} =\frac{1}{M} \sum_{i=1}^{P}  \sum_{u=0}^{M-1-l_{\tau_{i}}}\bigg[\overline{\gamma}_{n}^{i}(u+l_{\tau_{i}})  e^{-j 2 \pi\left((m-m^{\prime}) \Delta f-\nu_{i}\right)\left(\frac{u}{M\Delta f}+\tau_{i}\right)}\bigg]
	\end{aligned}
	\label{eq32}
	\end{equation}
	and ISI becomes
	\begin{equation}
	\begin{aligned}
	C_{nm,(n-1)m^{\prime}} =\frac{1}{M} \sum_{i=1}^{P}  \sum_{u=M-l_{\tau_{i}}}^{M-1}\bigg[\overline{\gamma}_{n-1}^{i}(u+l_{\tau_{i}}-T) \cdot e^{-j 2 \pi\left((m-m^{\prime}) \Delta f-\nu_{i}\right)\left(\frac{u}{M\Delta f}+\tau_{i}-T\right)}\bigg].
	\end{aligned}
	\label{eq33}
	\end{equation}
	Apparently, $\overline{\gamma}_{n-1}^{i}(u+l_{\tau_{i}}-T) = \overline{\gamma}_{n}^{i}(u+l_{\tau_{i}})$. Furthermore, we consider that one CP is added at the front of OTFS frame. Namely, the fragment of the $(N-1)$-th symbol period will enter into the first symbol period. Therefore, $C_{0m,-1m^{\prime}}$ makes sense and the TF relation can be simplified as 
	\begin{equation}
	Y_{nm} = \sum_{n^{\prime}=n-1}^{n} \sum_{m=0}^{M-1} C_{nm,n^{\prime} m^{\prime}} X_{n^{\prime} m^{\prime}} +W_{nm}.
	\label{eq33}
	\end{equation}
	We next characterize the delay-Doppler CIR and the input-output relation.
	\begin{myproposition}
		The delay-Doppler CIR with rectangular pulse can be characterized as
		\begin{equation}
		\begin{aligned}
		h_{kl,k^{\prime}l^{\prime}}^{\mathrm{DD}} = \frac{1}{N}\sum_{i=1}^{p}e^{j 2 \pi \left(\frac{l-l_{\tau_{i}}}{M}\right)\left(\frac{k_{\nu_{i}}+\kappa_{\nu_{i}}}{N}\right)}\alpha_{i}^{\mathrm{rect}}(k,k^{\prime},l) \delta([l-l^{\prime}-l_{\tau_{i}}]_{M})
		\end{aligned}
		\end{equation}
		and the demodulated signal $y_{kl}$
		\begin{equation}
		\begin{aligned}
		y_{kl} = \frac{1}{N}\sum_{i=1}^{p}e^{j 2 \pi \left(\frac{l-l_{\tau_{i}}}{M}\right)\left(\frac{k_{\nu_{i}}+\kappa_{\nu_{i}}}{N}\right)}\sum_{k^{\prime}=0}^{N-1}\alpha_{i}^{\mathrm{rect}}(k,k^{\prime},l) x\left[k^{\prime},[l-l_{\tau_{i}}]_{M}\right] +w_{kl}.
		\end{aligned}
		\label{eq36}
		\end{equation}
		where
		\begin{equation}
		\begin{aligned}
		\alpha_{i}^{\mathrm{rect}}(k,k^{\prime},l)=
		\begin{dcases}
		\alpha_{i}^{\mathrm{ideal}}(k,k^{\prime},l) e^{-j2\pi \frac{k^{\prime}}{N}} \: &l < l_{\tau_{i}}\\ 
		\alpha_{i}^{\mathrm{ideal}}(k,k^{\prime},l)  &l \geq l_{\tau_{i}}
		\end{dcases} 
		\end{aligned}
		\end{equation}
	\end{myproposition}
	\begin{myproof}
		See Appendix C.
	\end{myproof}
	Although the time-variant fading impacts on both ICI and ISI, the delay response has not been influenced. Analogously, rectangular pulse produces the identical property as ideal pulse, which is that $h^{\mathrm{DD}}$ can be interpreted as column-wise circular convolution of the original Doppler response with the discrete spectrum of $\boldsymbol{\gamma}_{i,l}$. For both pulses, $\gamma^{i}(t)$ will corrupt the dispersion in Doppler domain as follow
	\begin{figure*}[t]
		\centering
		\subfigure[]{
			\begin{minipage}[t]{0.32\linewidth}
				\centering
				\includegraphics[width=2.5in]{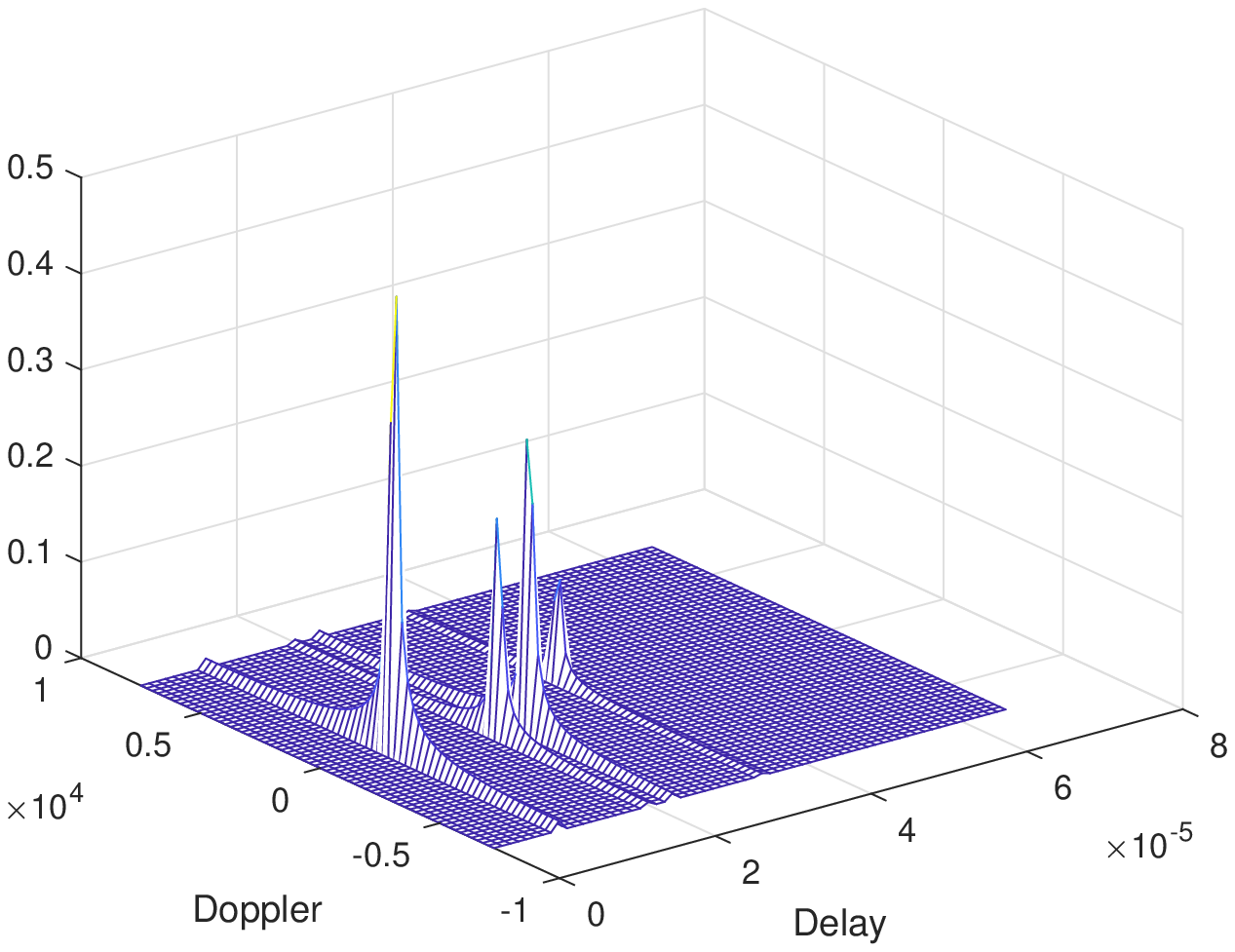}
			\end{minipage}%
		}%
		\subfigure[]{
			\begin{minipage}[t]{0.32\linewidth}
				\centering
				\includegraphics[width=2.5in]{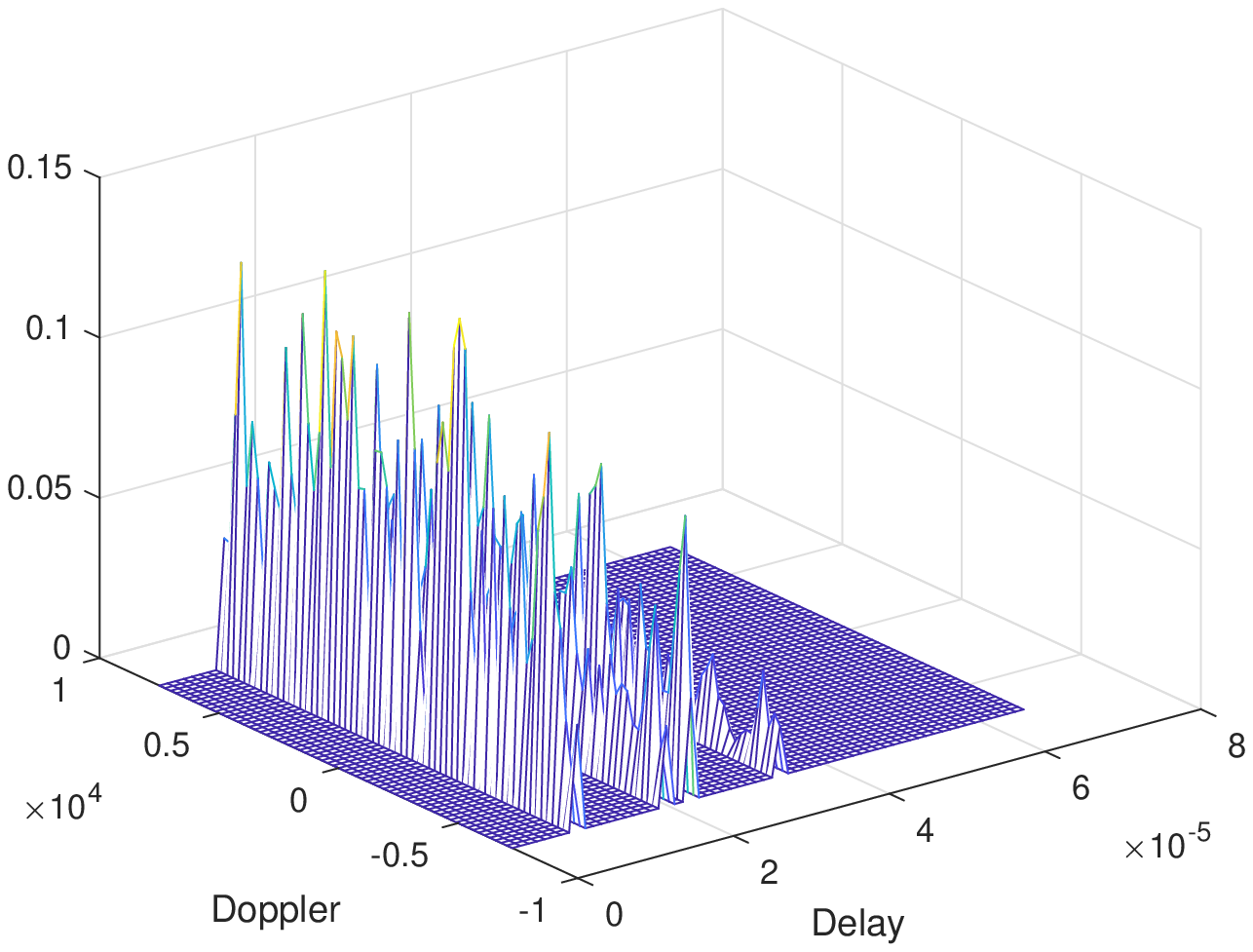}
			\end{minipage}%
		}%
		\subfigure[]{
			\begin{minipage}[t]{0.32\linewidth}
				\centering
				\includegraphics[width=2.5in]{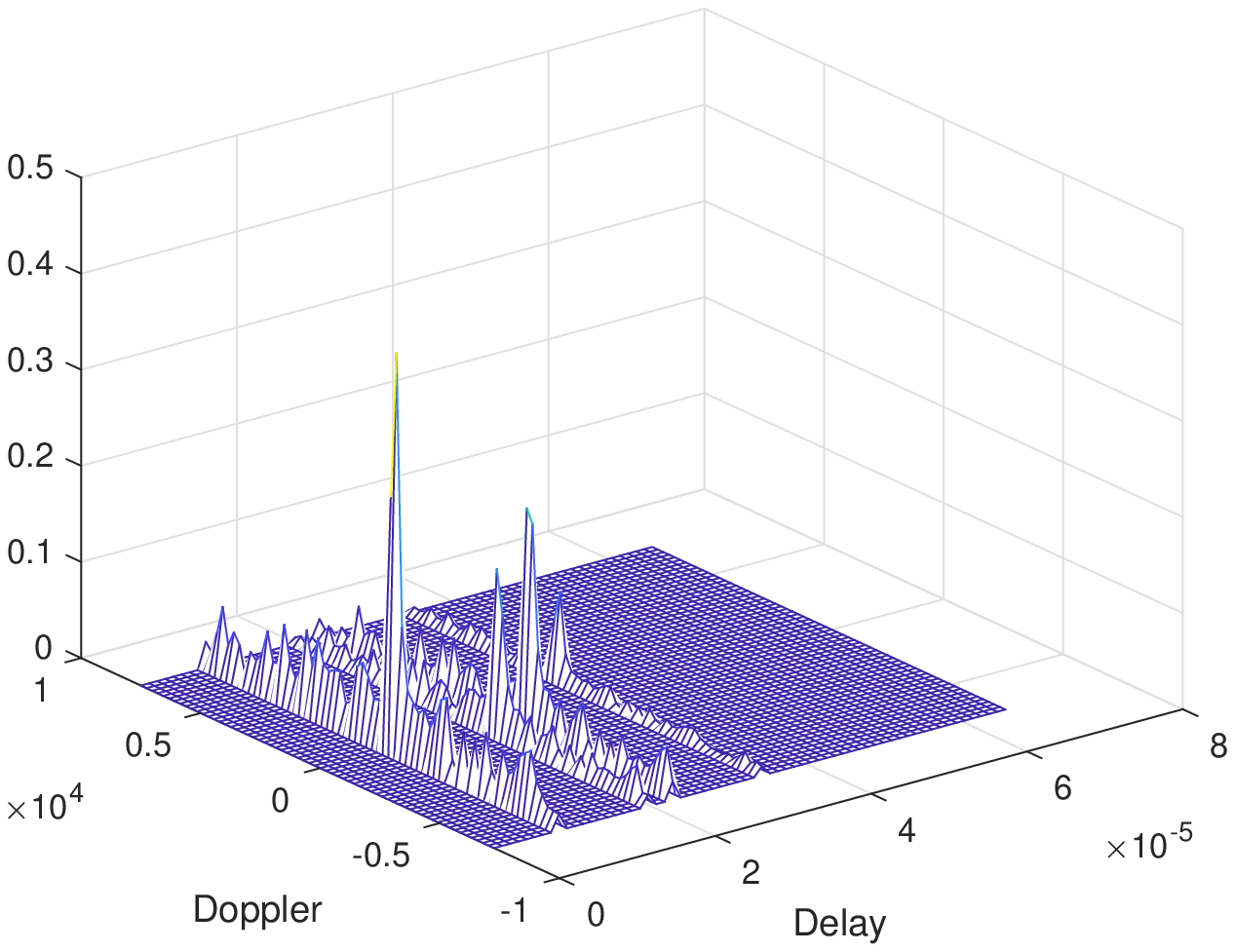}
			\end{minipage}
		}%
		\centering
		\caption{The CIR of different channels in DD domain. (a) quasi-static channel, (b) the worst case of rapid fading, where $\gamma_{i} \sim P^{-0.5} \mathcal{C} \mathcal{N}(0,1)$, (c) the ideal case of rapid fading, $\gamma_{i} \sim P^{-0.5}\mathcal{C}\mathcal{N}(0.8,0.36)$.}
		\label{Iter_FFT}
	\end{figure*}
	
	\begin{itemize}
		\item \emph{The worst case:} If $\gamma^{i}(t)$ is $\mathcal{C} \mathcal{N}(0,1)$ Rayleigh random function, the discrete amplitude spectrum of $\boldsymbol{\gamma}_{i,l}$ is approximately a constant. In this case, each symbol will uniformly disperse into other points of Doppler lattice. So with respect to the approximate ML detector, it is hard to perform perfect interference cancellation and result in degeneration of BER performance. Moreover, since the sparsity of Doppler domain has been broken, we cannot use the truncated Doppler response in detector.
		\item \emph{The ideal case:} If the energy of $\gamma^{i}(t)$ is well gathered around $N$ times frequency of $\Delta f/N$, where $N$ is an arbitrary positive integer, and tends to be 0 with increasing frequency, the extra Doppler interference is finite as well. For the non-zero fractional Doppler, since it has been proven that the original Doppler response has a peak around $k^{\prime}=k-k_{\nu_{i}}$ and decays rapidly as $k^{\prime}$ moving away from $k-k_{\nu_{i}}$, the convolution result will still decrease with a more moderate slope. Hence, (\ref{eq29}) can be expressed as a sparse linear system as before. Despite $\gamma^{i}(t)$ produces additional elements in CIR, the complexity of detector will not increase by using truncated Doppler response, for the reason that the connection of the corresponding probability graph has not been changed. 
	\end{itemize}
	
	Intuitively, since it is difficult to eliminate more dispersed Doppler interference, the BER for rapid fading channels will be higher than that for quasi-static channels. However, simulation results show the opposite conclusion. Notice that the motivation behind OTFS is to spread delay-Doppler domain symbol to time domain, so that each symbol has experienced more channel states. We speculate that OTFS involves inherent time diversity.
	
	\section{Time Diversity Analysis}
	
	In this section, we will show that OTFS can be considered as a kind of technique that combines encoding and interleaving to achieve time diversity. This means that QAM symbols in delay-Doppler domain becomes insensitive to rapid fading when the $N$ is large. Firstly, we commence from the vectorized form of input-output relation in OTFS. The received signal $\mathbf{y}$ of size $NM \times 1$ can be rewritten as
	\begin{equation}
	\mathbf{y}=\underbrace{\left( \mathbf{F}_{N} \otimes \mathbf{I}_{M}\right) \mathbf{H} \left( \mathbf{F}_{N}^{\rm{H}} \otimes \mathbf{I}_{M}\right)}_{\mathbf{H}^\text{DD}}\mathbf{x}+ \mathbf{w}.
	\end{equation}
	$\otimes$ denotes Kronecker product, $\mathbf{x}$ of size $NM \times 1$ is column-wise rearranged data symbols, $\mathbf{F}_{N}$ is the $N$-points DFT matrix, and $\mathbf{I}_{M}$ is the $M \times M$ identity matrix. From the interaction between delay-Doppler symbols and channel, the channel matrix $\mathbf{H}$ is given by
	\begin{equation}
	\mathbf{H} =\sum_{i=1}^{P}  \mathbf{\Gamma}^{i} \mathbf{\Delta}^{\left(k_{\nu_{i}}+\kappa_{\nu_{i}}\right)} \mathbf{\Pi}^{l_{\tau_{i}}},
	\end{equation}
	where $\mathbf{\Gamma}^{i}=\operatorname{diag}\left[\boldsymbol{\gamma}_{i,0}^{T},\boldsymbol{\gamma}_{i,1}^{T},\cdots , \boldsymbol{\gamma}_{i,M-1}^{T}\right]$ is an $NM \times NM$ fading diagonal matrix, $ \mathbf{\Delta}$ of size $NM \times NM$ denotes the diagonal Doppler matrix $\operatorname{diag}\left[z_{0}^{\rm{T}}, z_{1}^{\rm{T}},\cdots, z_{M-1}^{\rm{T}}\right]$, in which $z_{m}=\left[e^{j2\pi \frac{mN}{NM}},e^{j2\pi \frac{mN+1}{NM}}, \cdots, e^{j2\pi \frac{(m+1)N-1}{NM}}\right]^{\rm{T}}$, and $\mathbf{\Pi}$ is the $NM \times NM$ block circulant delay matrix	
	\begin{equation}
	\mathbf{\Pi}_{n}=
	\begin{bmatrix}
	\bm{0} & \cdots & \bm{0} & \mathbf{I}_{M} \\
	\mathbf{I}_{M} & \cdots & \bm{0} & \bm{0} \\
	\vdots & \ddots & \vdots & \vdots \\
	\bm{0} & \cdots & \mathbf{I}_{M} & \bm{0}
	\end{bmatrix}
	\end{equation}
	
	We focus the proof on the pairwise error probability (PEP) $P\left(\mathbf{x} \rightarrow \widetilde{\mathbf{x}} \right)$, which is the probability of transmitting $\mathbf{x}$ and deciding in favor of $\widetilde{\mathbf{x}}$ at receiver. Assuming perfect CSI and ML detection are available, the conditional PEP is given by
	\begin{equation}
	P(\mathbf{x} \rightarrow \widetilde{\mathbf{x}} \mid \mathbf{H}^\text{DD}) = P
	\left(\|\mathbf{y}-\mathbf{H}^\text{DD}\widetilde{\mathbf{x}} \|^{2}<\|\mathbf{y}-\mathbf{H}^\text{DD}\mathbf{x} \|^{2}\right)
	\end{equation}
	Let us assume that the average energy of constellation elements is $1$ and the noise $\mathbf{w}_{i}$ is zero mean, $N_{0}$ variance Gaussian distributed independent random variables, the conditional PEP can be written as
	\begin{equation}
	P(\mathbf{x} \rightarrow \widetilde{\mathbf{x}} \mid \mathbf{H}^\text{DD}) =Q\left(\sqrt{\frac{\left\|\mathbf{H}^\text{DD}\left(\mathbf{x}-\widetilde{\mathbf{x}}\right)\right\|^{2}}{2 N_{0}}}\right),
	\end{equation}
	where $Q$ denotes Gaussian tail function. We regard $\left( \mathbf{F}_{N}^{\rm{H}} \otimes \mathbf{I}_{M}\right)$ and $\mathbf{\Pi}^{l_{\tau_{i}}}$ as coding matrices, and the codeword $\bm{c}^{i}=\mathbf{\Pi}^{l_{\tau_{i}}}\left( \mathbf{F}_{N}^{\rm{H}} \otimes \mathbf{I}_{M}\right)\mathbf{x}$, whose the $(mN+n)$-th entry is denoted as $\bm{c}^{i}_{n,m}$. The $\left\|\mathbf{H}^\text{DD}\left(\mathbf{x}-\widetilde{\mathbf{x}}\right)\right\|^{2}$ is calculated as
	\begin{equation}
	\begin{aligned}
	\left\|\mathbf{H}^\text{DD}\left(\mathbf{x}-\widetilde{\mathbf{x}}\right)\right\|^{2}  &= \left(\mathbf{x}-\widetilde{\mathbf{x}}\right)^{\rm{H}} \left( \mathbf{F}_{N} \otimes \mathbf{I}_{M}\right)\mathbf{H}^{\rm{H}}\mathbf{H} \left( \mathbf{F}_{N}^{\rm{H}} \otimes \mathbf{I}_{M}\right)\left(\mathbf{x}-\widetilde{\mathbf{x}}\right) \\
	&=\sum_{n,m}^{N,M}\left|\sum_{i=1}^{P}  \overline{\gamma}_{n}^{i}(m) z_{m}^{k_{\nu_{i}}+\kappa_{\nu_{i}}}(n) (\bm{c}^{i}_{n,m}-\widetilde{\bm{c}}^{i}_{n,m})\right|^{2} \\
	&=\sum_{n,m}^{N,M} \mathbf{\Omega}_{n,m} \mathbf{C}_{n,m} \mathbf{\Omega}^{\rm{H}}_{n,m}.
	\label{eq43}
	\end{aligned}
	\end{equation}
	In (\ref{eq43}), the $\mathbf{C}_{n,m}$ is an $P \times P$ matrix
	\begin{equation}
	\mathbf{C}_{n,m}=
	\begin{bmatrix}
	\left|\bm{c}^{1}_{n,m}-\widetilde{\bm{c}}^{1}_{n,m}\right|^{2}  & \left(\bm{c}^{1}_{n,m}-\widetilde{\bm{c}}^{1}_{n,m}\right)\left(\bm{c}^{2}_{n,m}-\widetilde{\bm{c}}^{2}_{n,m}\right)& \cdots & \left(\bm{c}^{1}_{n,m}-\widetilde{\bm{c}}^{1}_{n,m}\right)\left(\bm{c}^{P}_{n,m}-\widetilde{\bm{c}}^{P}_{n,m}\right) \\
	\left(\bm{c}^{2}_{n,m}-\widetilde{\bm{c}}^{2}_{n,m}\right)\left(\bm{c}^{1}_{n,m}-\widetilde{\bm{c}}^{1}_{n,m}\right) & \left|\bm{c}^{2}_{n,m}-\widetilde{\bm{c}}^{2}_{n,m}\right|^{2} & \cdots & \left(\bm{c}^{2}_{n,m}-\widetilde{\bm{c}}^{2}_{n,m}\right)\left(\bm{c}^{P}_{n,m}-\widetilde{\bm{c}}^{P}_{n,m}\right) \\
	\vdots &\vdots  & \ddots & \vdots \\
	\left(\bm{c}^{P}_{n,m}-\widetilde{\bm{c}}^{P}_{n,m}\right)\left(\bm{c}^{1}_{n,m}-\widetilde{\bm{c}}^{1}_{n,m}\right) & \left(\bm{c}^{P}_{n,m}-\widetilde{\bm{c}}^{P}_{n,m}\right)\left(\bm{c}^{2}_{n,m}-\widetilde{\bm{c}}^{2}_{n,m}\right) & \cdots & \left|\bm{c}^{P}_{n,m}-\widetilde{\bm{c}}^{P}_{n,m}\right|^{2},
	\end{bmatrix}
	\label{eq44}
	\end{equation} 
	and $\Omega_{n,m}$ is given by 
	\begin{equation}
	\mathbf{\Omega}_{n,m} = \left[\overline{\gamma}_{n}^{1}(m) z_{m}^{k_{\nu_{1}}+\kappa_{\nu_{1}}}(n),  \overline{\gamma}_{n}^{2}(m) z_{m}^{k_{\nu_{2}}+\kappa_{\nu_{2}}}(n), \cdots, \overline{\gamma}_{n}^{P}(m) z_{m}^{k_{\nu_{P}}+\kappa_{\nu_{P}}}(n)\right]
	\label{eq45}
	\end{equation}
	
	The matrix $\mathbf{C}_{n,m}$ is Hermitian, thus there exists eigenvalue decomposition such that $\mathbf{C}_{n,m}= \mathbf{U}_{n,m}\mathbf{\Lambda}_{n,m}\mathbf{U}_{n,m}^{\rm{H}}$. $\mathbf{U}_{n,m}$ is unitary matrix and $\mathbf{\Lambda}_{n,m}=\operatorname{diag}\left[\lambda_{n,m}^{1},\lambda_{n,m}^{2},\cdots,\lambda_{n,m}^{P}\right]$. Let
	\begin{equation}
	\left[\eta_{n,m}^{1},\eta_{n,m}^{2},\cdots,\eta_{n,m}^{P}\right] = \mathbf{\Omega}_{n,m} \mathbf{U}_{n,m},
	\end{equation}
	then it yields that 
	\begin{equation}
	\left\|\mathbf{H}^\text{DD}\left(\mathbf{x}-\widetilde{\mathbf{x}}\right)\right\|^{2} = \sum_{n,m}^{N,M}\sum_{i=1}^{P} \left|\eta_{n,m}^{i}\right|^{2}\lambda_{n,m}^{i}.
	\end{equation}
	By utilizing Chernoff bound technique, we have the following average PEP
	\begin{equation}
	\begin{aligned}
	P(\mathbf{x} \rightarrow \widetilde{\mathbf{x}}) &= \mathbb{E} \left[P(\mathbf{x} \rightarrow \widetilde{\mathbf{x}} \mid \mathbf{H}^\text{DD})\right] \\
	&\leq \mathbb{E} \left[\operatorname{exp}\left(-\dfrac{\sum\limits_{n,m}^{N,M}\sum\limits_{i=1}^{P} \left|\eta_{n,m}^{i}\right|^{2}\lambda_{n,m}^{i}}{4N_{0}}\right)\right]
	\label{eq46}
	\end{aligned}
	\end{equation}
	Suppose that $\overline{\gamma}_{n}^{i}(m)$ for $i=1,2,\cdots,P$, $n=0,1,\cdots,N-1$, $m=0,1,\cdots,M-1$ are samples of independent zero-mean complex Gaussian random variables with variance $1$. Since $\mathbf{U}_{n,m}$ is unitary, $\eta_{n,m}^{i}$ follows the same distribution $\mathcal{C}\mathcal{N}(0,1)$. Hence, we have that
	\begin{equation}
	P(\mathbf{x} \rightarrow \widetilde{\mathbf{x}}) \leq \prod_{i,n,m} \dfrac{1}{1+\dfrac{\lambda_{n,m}^{i}}{4N_{0}}}
	\label{eq47}
	\end{equation}
	Based on the upper bound on average PEP, the diversity advantage is the number of non-zero $\lambda_{n,m}^{i}$. It is obvious that the rows of $\mathbf{C}_{n,m}$ are all linearly dependent. Thus $\mathbf{C}_{n,m}$ has rank $1$ if $\bm{c}_{n,m}=\left[\bm{c}^{1}_{n,m},\bm{c}^{2}_{n,m},\cdots,\bm{c}^{P}_{n,m}\right]$ is distinct from $\widetilde{\bm{c}}_{n,m}=\left[\widetilde{\bm{c}}^{1}_{n,m},\widetilde{\bm{c}}^{2}_{n,m},\cdots,\widetilde{\bm{c}}^{P}_{n,m}\right]$, and the non-zero eigenvalue is $\left|\bm{c}_{n,m}-\widetilde{\bm{c}}_{n,m}\right|^{2}$. Otherwise, the rank is $0$. Let $\mathcal{V}(\bm{c}, \widetilde{\bm{c}})$ donates the set of indexes with $\mathbf{c}_{n,m} \neq \widetilde{\bm{c}}_{n,m}$, at high SNR, (\ref{eq47}) can be further simplified as
	\begin{equation}
	P(\mathbf{x} \rightarrow \widetilde{\mathbf{x}}) \leq \prod_{n,m \in \mathcal{V}(\bm{c}, \widetilde{\bm{c}})} \left|\bm{c}_{n,m}-\widetilde{\bm{c}}_{n,m}\right|^{2} \left(\frac{1}{4N_{0}}\right)^{-r},
	\label{eq48}
	\end{equation}
	where $r$ is the number of elements in $\mathcal{V}(\mathbf{c}, \widetilde{\bm{c}})$. (\ref{eq48}) reveals that OTFS modulation and multipath delay provide both coding gain (the $r$-product distance) and diversity gain. Since $\bm{c}^{i}$ for $i=2,3,\cdots,P$ are circular shift of $\bm{c}^{1}$, the optimal situation is that all the different elements between $\bm{c}^{1}$ and $\widetilde{\bm{c}}^{1}$ are moved to a unique position. Therefore, the maximum achievable diversity is $\operatorname{min}\left(Pd(\bm{c}^{1},\widetilde{\bm{c}}^{1}),NM\right)$, where $d(\bm{c}^{1},\widetilde{\bm{c}}^{1})$ is the Hamming distance between $\bm{c}^{1}$ and $\widetilde{\bm{c}}^{1}$.
	
	\begin{figure}[!t]
		\centering
		\includegraphics[width=3.5in]{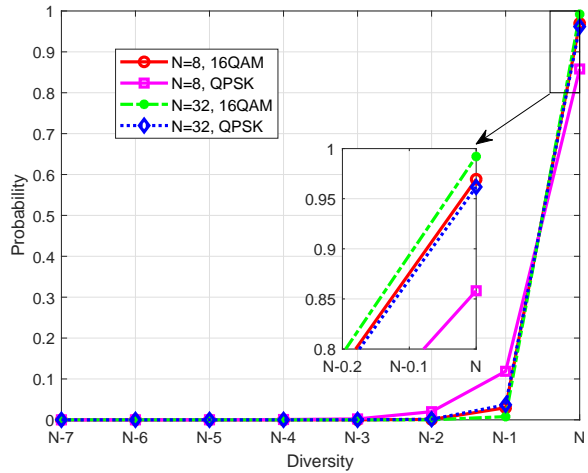}
		\caption{Diversity distribution for DFT matrix for different size $N$ and modulation order.}
		\label{fig1}
	\end{figure}
	
	We first investigate $d(\bm{c}^{1},\widetilde{\bm{c}}^{1})$. $N$-points IDFT matrix encodes every $N$ data symbols in $\mathbf{x}$ into sub-codewords, and $M$ subcodewords together compose the codeword $\bm{c}^{1}$, hence the IDFT has a direct impact on the diversity order. Now, consider two column vectors $\bm{a}$ and $\widetilde{\bm{a}}$ of size $N$. When $\bm{a}\left[i\right]=a$ and $\bm{a}\left[i\right]=a^{\prime}$, $\forall i=0,1,\cdots,N-1$, the IDFT result of difference vector $\bm{a}-\widetilde{\bm{a}}$ will has $N$ non-zero elements. On the contrary, when $\bm{a}\left[i\right]=a\bm{1}_{N\times 1}$ and $\bm{a}\left[i\right]=a^{\prime}\bm{1}_{N\times 1}$, the Hamming distance will be $1$. So, in the strict sense of definition, IDFT matrix cannot provide any diversity gain. However, However, the Hamming distance of most pairwise codewords is $N$ in practice. The reason is that only if $\bm{a}-\widetilde{\bm{a}}$ is linearly dependent on any column vector of $N$-point DFT matrix, the Hamming distance will be $1$. Here, diversity distribution is introduced to explain this result \cite{850699}. Since an analytical formula for the diversity distribution of IDFT matrix could be intractable to evaluate, we carry out simulation in Matlab. As illustrated in Fig. \ref{fig1}, large $N$ and high order QAM will provide full diversity advantage. Therefore, the minimum $d(\bm{c}^{1},\widetilde{\bm{c}}^{1}) \approx N$. Considering that these $N$ entries will inevitably appear in new positions after being shifted, the minimum achievable $r$ for any pair of distinct codewords is $PN$. From the above analysis, (\ref{eq48}) becomes
	\begin{equation}
	P(\mathbf{x} \rightarrow \widetilde{\mathbf{x}}) \leq \prod_{n,m \in \mathcal{V}(\bm{c}, \widetilde{\bm{c}})} \left|\bm{c}_{n,m}-\widetilde{\bm{c}}_{n,m}\right|^{2} \left(\frac{1}{4N_{0}}\right)^{-PN},
	\label{eq49}
	\end{equation}
	The $P$ is also known as delay diversity advantage \cite{140615}.
	
	In multipath OTFS system, the maximum available degrees of freedom in the channel is upper-bounded by $2NM$, where the term $2$ represents real part and imaginary part. As for time-correlated rapid fading Channels, the maximum achievable diversity will be less than $2NM$ \cite{1317131}. Apparently, the uncoded OTFS can not achieve the maximum time diversity order. The full diversity order can be available by utilizing signal space diversity (SSD) technology \cite{485720,681321}. Nevertheless, simulation results show that it is unnecessary to increase the computation complexity in exchange for a weak BER performance improvement for practical OTFS system.
	
	\section{Simulation Results}
	
	In this section, we illustrate the performance in terms of BER of OTFS over rapid fading channels to reveal that OTFS modulation involves inherent time diversity gain. A carrier frequency of 4 GHz and a carrier separation of 15 kHz are considered. The channel gains are assumed to be i.i.d and distributed as $\mathcal{C} \mathcal{N}(0,1 / P)$. For each path, the delay index is random integer with equal probabilities from the set $\{0,1,\cdots,M-1\}$ and the Doppler index is randomly generated belongs to $[0,N-1]$. 
	
	\begin{figure*}[!t]

		\centering
		\includegraphics[width=5.5in]{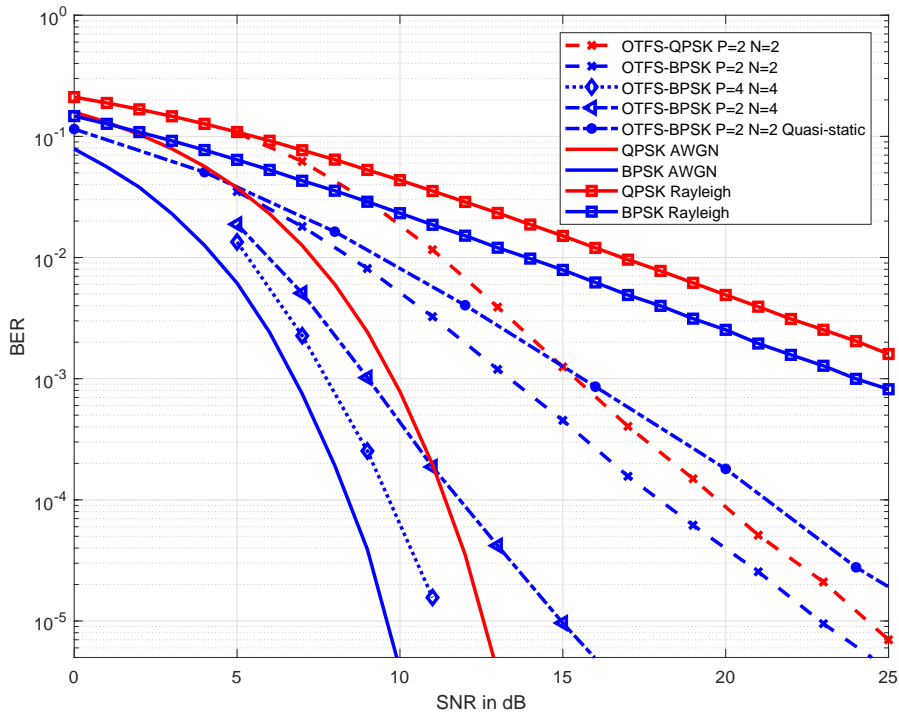}
		\caption{The BER performance comparison for different paths, Doppler lattice and modulation order with ML detector.}
		\label{fig3}
	\end{figure*}
	
	\begin{figure}[!t]
		\centering
		\includegraphics[width=3.5in]{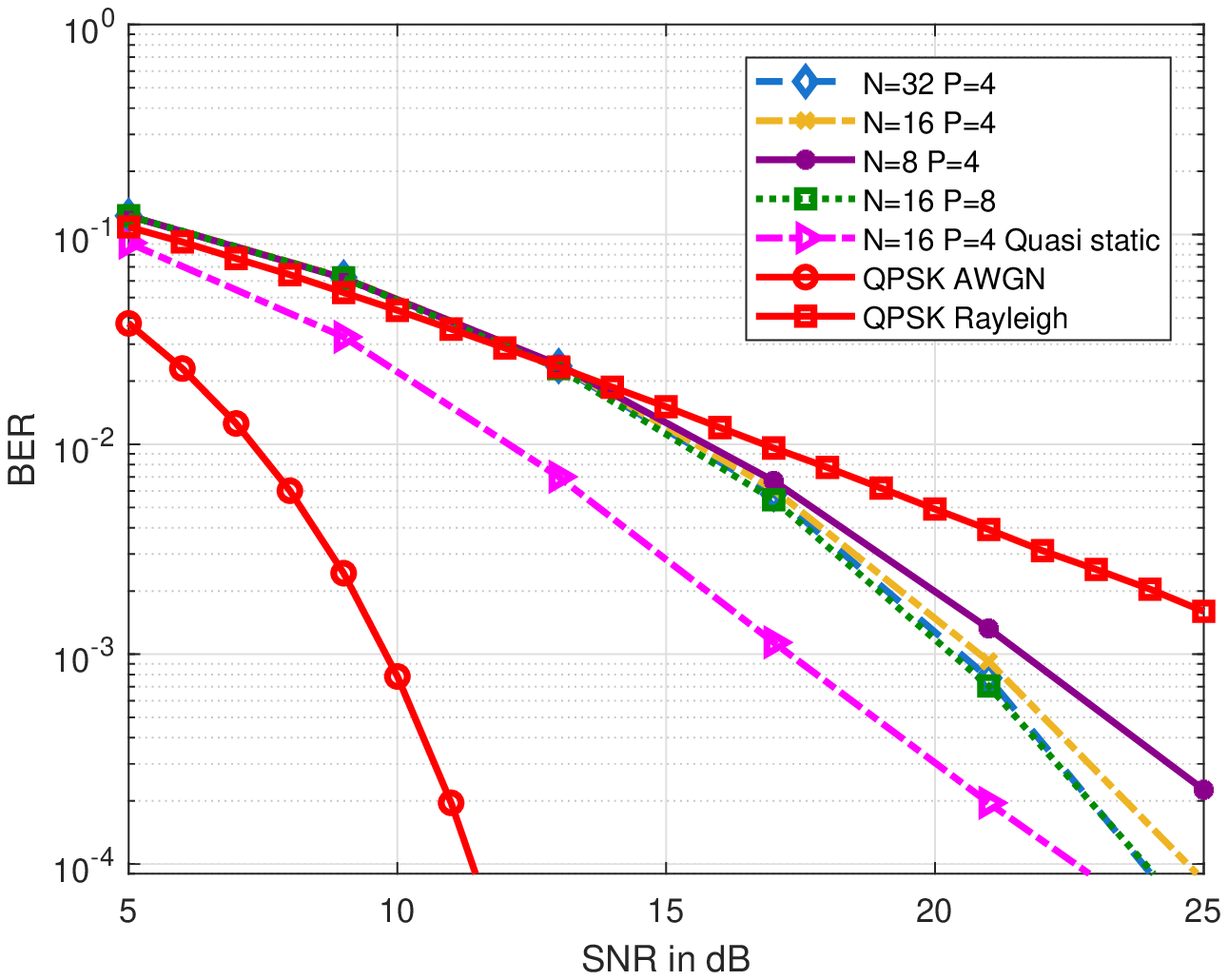}
		\caption{The BER performance comparison for different paths, Doppler lattice with MMSE detector.}
		\label{fig4}
	\end{figure}
	
	\begin{figure}[!t]
		\centering
		\includegraphics[width=3.5in]{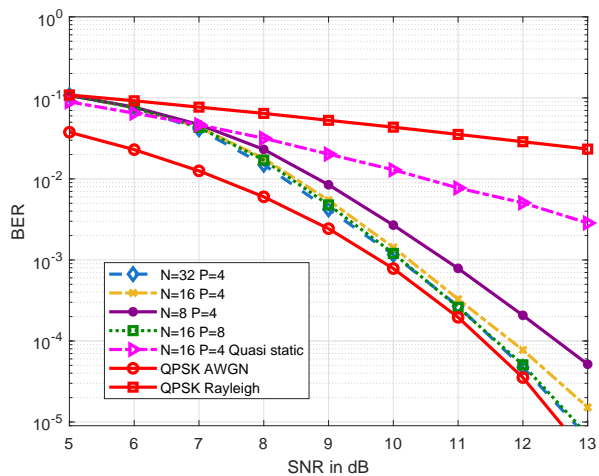}
		\caption{The BER performance comparison for different paths, Doppler lattice with MP detector.}
		\label{fig5}
	\end{figure}
	
	First, we simulate the BER performance with the ML detector. Since the ML detector has a complexity exponential in $NM$, we study the case of small values of $M$ and $N$ to illustrate the time diversity order of OTFS modulation. Figure \ref{fig3} shows the BER performance in various scenarios. We plot the BER curves of BPSK and QPSK over the AWGM channel and over the rapid Rayleigh fading channel, so the time diversity gain of OTFS is bounded by these two curves. For low SNR, since the term 1 in (\ref{eq47}) is the dominating factor affecting PEP and can not be ignored, the BER performance of OTFS is the same as that of Rayleigh fading channel. As SNR increases, OTFS can obtain the diversity gain. The family of blue BER curves of OTFS-BPSK in different paths and Doppler lattices demonstrates that larger $P$ and $N$ can yield higher diversity order. It turns out that the BER curves will approach the one for the Gaussian channel as $N$ and $P$ increase, and the gap between the maximum diversity value and the Gaussian BER curve is only about 2 dB around $10^{-5}$. Furthermore, it can be observed that the gap between OTFS-BPSK and OTFS-QPSK gradually reduces when the SNR increases, since high-order QAMs can bring asymptotic full diversity order. It is also worth noting that the BER performance of OTFS over rapid fading channels is superior to that of quasi-static multipath channels. The reason is that the symbols in the DD domain have experienced more various channel states. so that the fading coefficients can be sufficiently averaged over time. As a result, the OTFS system becomes insensitive to rapid fading.
	
	Next, we considered more practical values of $N$ and $M$ with MMSE and MP detector. In Figure \ref{fig4} and Figure \ref{fig5}, we present the BER performance of OTFS system with $M=16$ and $N=8,16,32$, respectively. The maximum speed of user velocity is set to be 500 km/h, which corresponds to a Doppler of 1.85 kHz at 4 GHz carrier frequency. For the $i$th tap, Doppler shift is generated using $v_{i}=v_{\operatorname{max}}\operatorname{cos}\left(\theta_{i}\right)$, where $v_{\operatorname{max}}$ is the maximum Doppler shift and $\theta_{i} \sim \mathcal{U}(0, \pi)$ is uniformly distributed.
	
	\begin{figure}[!t]
		\centering
		\includegraphics[width=3.5in]{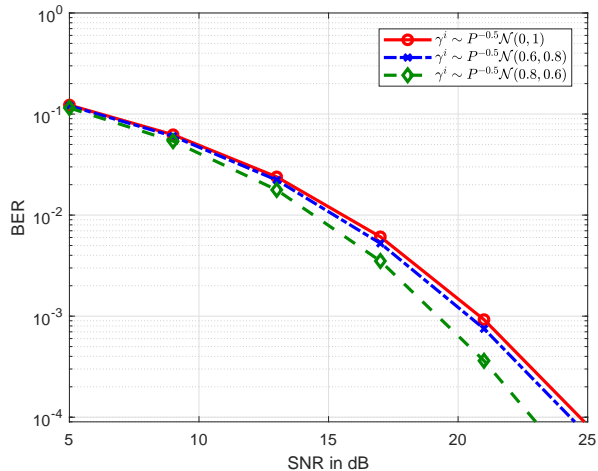}
		\caption{The BER performance comparison for different distribution of $\gamma^{i}$ with MMSE detector.}
		\label{fig6}
	\end{figure}
	
	\begin{figure}[!t]
		\centering
		\includegraphics[width=3.5in]{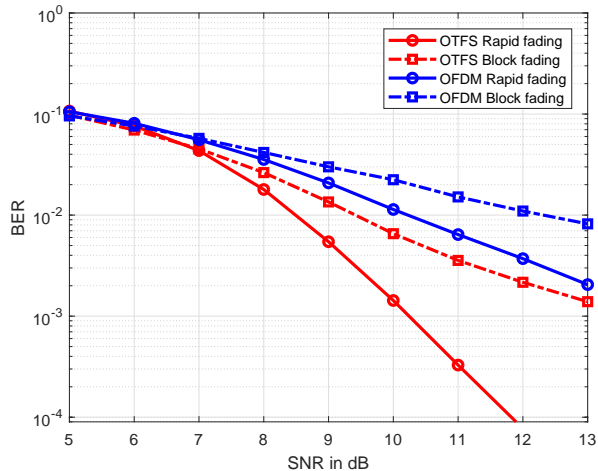}
		\caption{The BER performance comparison between OTFS and OFDM, $M=N=16$, $P=4$.}
		\label{fig7}
	\end{figure}
	
	In Figure \ref{fig4} and Figure \ref{fig5}, we simulate MMSE and MP detector, respectively. It can be seen that the BER will decrease with the increasing $N$ and $P$, where the BER performance of the case of $N=32$, $P=4$ is almost the same as that of the case of $N=16$, $P=8$. This result confirms the correctness of the diversity order that we proved. Compared with MMSE and MP detector, it can be seen that the MMSE detector suffers a significant diversity loss. The MMSE detector can obtain diversity gain as SNR exceeds 15 dB, and the gap between MP detector and MMSE detector is more than 20 dB around $10^{-4}$. it is noticed that the BER performance of MMSE detector for rapid fading channel is worse than that for quasi-static channel, which distinguishes from ML and MP detector. The reason is that the BER performance of MMSE detector is extremely dependent on the condition number of $\mathbf{H}^\text{DD}$ . As we analysed in Section III, the Rayleigh rapid fading will disperse the CIR of Doppler axis and make $\mathbf{H}^\text{DD}$ tend to be ill-condition. To illustrate this, the BER performance for various distributions of $\gamma^{i}$ with the same power is plotted in Figure \ref{fig6}. The larger mean value of $\gamma^{i}$ makes CIR more centralized in the Doppler domain, making a smaller condition number of $\mathbf{H}^\text{DD}$.
	
	In Figure \ref{fig7}, we compare the BER performance between OTFS and OFDM with the same TF resources. Since OFDM is encoded by the IDFT matrix, it can also achieve part time diversity gain in rapid channels. However, OFDM symbols have only experienced the same number of channel states as that of subcarriers, so the delay diversity gains can not be obtained. In the block fading channel, OFDM frame has only experienced one channel state and cannot obtain any diversity gain, while OTFS will only lose the delay diversity gain. Consequently, OTFS still outperforms OFDM in multipath rapid fading channels.
	
	\section{Conclusions}
	
	In this paper, we have investigated the characterizations of OTFS modulation over
	multipath rapid fading channel. We have analyzed the interaction between WH basis and the above channel, and derived input-output relation in DD domain for the cases of ideal pulse and rectangular pulse. Despite the rapid fading will degenerate both ISI and ICI, the impact on CIR of domain is limited. We have shown that the sparsity in delay domain has not been destroyed, while the response in Doppler domain can be formulated as a column-wise circular convolution of the original Doppler response with the extra Doppler dispersion. 
	Then, we have proven that OTFS is a time diversity technique which can achieve $PN$ diversity order. For large $P$ and $N$, the BER performance is almost consistent with Gaussian performance over the rapid fading channels. Therefore, OTFS is still a reliable communication scheme in most practical multipath channel. To achieve the time diversity gain, the perfect CSI should be available at receiver. In particular, the parameters to be estimated increase manyfold for the case of Rayleigh rapid fading. Designing an efficient estimation algorithm can be considered for future work. 
	
	\appendices
	\section{Proof of Theorem 1}
	From (\ref{eq24}), the dispersion introduced by $\gamma(t)$ among WH basis is written as
	\begin{equation}
	\begin{aligned}
	\langle \gamma(t)&g_{n^{\prime}m^{\prime}}(t), g_{nm}(t)\rangle \\
	&= 
	\biggl(\int_{t} g^{*}(t-(n-n^{\prime})T) g(t) \gamma(t+n^{\prime}T) e^{-j2\pi(m-m^{\prime})\Delta ft}dt \biggr) e^{j2\pi m\Delta f(n-n^{\prime})T}
	\end{aligned}
	\end{equation}
	At the receiver, the signal is sampled at intervals of $1/M\Delta f$ over duration $T$. Therefore, we have an approximation to the inner product
	\begin{equation}
	\begin{aligned}
	\langle \gamma(t)&g_{n^{\prime}m^{\prime}}(t), g_{nm}(t)\rangle = \frac{1}{M\Delta f} \Bigg[\sum_{u=0}^{M-1} \underbrace{\gamma\left(\frac{u}{M\Delta f}+n^{\prime}T\right)}_{\overline{\gamma}_{n}(u)} \\ & \quad \cdot \underbrace{g\left(\frac{u}{M\Delta f}\right) g^{*}\left(\frac{u}{M\Delta f}-(n-n^{\prime})T\right) e^{j2\pi \frac{m^{\prime}}{M}u}}_{\epsilon(u)} e^{-j2\pi \frac{u}{M}(m)}\Bigg]  e^{j2\pi m\Delta f(n-n^{\prime})T}.
	\end{aligned}
	\label{eqA1}
	\end{equation}
	Obviously, the item in square brackets can be interpreted as an $M$-points DFT of $\overline{\gamma}_{n}(u)\epsilon(u)$. Hence, the following relation satisfies
	\begin{equation}
	\text{DFT}(\overline{\gamma}_{n}(u)\epsilon(u))	= \frac{1}{M} \text{DFT}(\overline{\gamma}_{n}(u)) \circledast \text{DFT}(\epsilon(u)),
	\end{equation}
	Based on this property, we next calculate the DFT of ${\gamma}(u)$ and $\epsilon(u)$, respectively. Multiplying $\text{DFT}(\epsilon(u))$ by the coefficient outside the square brackets, we have that
	\begin{equation}
	\begin{aligned}
	\frac{1}{M\Delta f} \text{DFT}(\epsilon(u)) e^{j2\pi m\Delta f(n-n^{\prime})T}  &= \Bigg[\frac{1}{M\Delta f} \sum_{u=0}^{M-1}  g^{*}\left(\frac{u}{M\Delta f}-(n-n^{\prime})T\right) \\ & \quad \cdot g\left(\frac{u}{M\Delta f}\right) e^{-j2 \pi (m-m^{\prime})\Delta f\left(\frac{u}{M\Delta f}-(n-n^{\prime})T\right)}\Bigg] e^{j2 \pi m^{\prime} \Delta f (n-n^{\prime})T} \\
	&\approx \Bigg[\int_{t=0}^{T} g(t) g^{*}_{n-n^{\prime},m-m^{\prime}}(t)dt\Bigg] e^{j2 \pi m^{\prime} \Delta f (n-n^{\prime})T} \\
	&= \delta (n-n^{\prime},m-m^{\prime}).
	\label{eqA2}
	\end{aligned}
	\end{equation}
	In (\ref{eqA2}), the item on the right side of the first equal sign can be regarded as a discretization approximation to evaluate the $\langle g(t), g_{n-n^{\prime},m-m^{\prime}}(t) \rangle$ within duration $T$. Generally, $g(t)$ has well-defined energy in time interval $\left[0,T\right]$, which makes the orthogonality of the WH basis assumed to be held. Combined with the result in (\ref{eqA2}), it immediately yields that
	\begin{equation}
	\begin{aligned}
	\langle \gamma(t)g_{n^{\prime}m^{\prime}}(t), g_{nm}(t)\rangle
	= 
	\begin{dcases}
	\frac{1}{M}\sum\limits_{u=0}^{M-1} \overline{\gamma}_{n}(u) e^{-j2\pi\frac{u}{M}(m-m^{\prime})} \quad &n=n^{\prime}\\ 
	0 \quad &n\neq n^{\prime}
	\end{dcases} 
	\end{aligned}
	\end{equation}
	
	\section{Proof of Proposition 1}
	By (\ref{eq16}), (\ref{eq26}) and (\ref{eq27}), the output signal $y_{kl}$ is given by
	\begin{equation}
	\begin{aligned}
	y_{kl}=&\frac{1}{N M} \sum_{n=0}^{N-1} \sum_{m=0}^{M-1} \sum_{m^{\prime}=0}^{M-1} C_{nm^{\prime},nm^{\prime}}  \sum_{k^{\prime}=0}^{N-1} \sum_{l^{\prime}=0}^{M-1} x_{k^{\prime}l^{\prime}} e^{j 2 \pi\left(\frac{n k^{\prime}}{N}-\frac{m^{\prime} l^{\prime}}{M}\right)} e^{-j 2 \pi\left(\frac{n k}{N}-\frac{m l}{M}\right)} \\
	=&  \frac{1}{N M}\sum_{k^{\prime}=0}^{N-1} \sum_{l^{\prime}=0}^{M-1} x_{k^{\prime}l^{\prime} } \Bigg[ \sum_{n=0}^{N-1} \sum_{m=0}^{M-1} \sum_{m^{\prime}=0}^{M-1} \frac{1}{M} \sum_{i=1}^{p} H^{i}_{nm^{\prime},nm^{\prime}}\sum_{u=0}^{M-1} \overline{\gamma}_{n}^{i}(u) e^{-j2\pi\frac{u}{M}(m-m^{\prime})} \\ & \quad e^{-j2\pi n\left(\frac{k-k^{\prime}}{N}\right)} e^{j2\pi \left(\frac{ml-m^{\prime}l^{\prime}}{M}\right)} \Bigg]\\
	=&  \frac{1}{N M}\sum_{k^{\prime}=0}^{N-1} \sum_{l^{\prime}=0}^{M-1} x_{k^{\prime}l^{\prime} } h^{\mathrm{DD}}_{kl,k^{\prime}l^{\prime}}.
	\end{aligned}
	\end{equation}
	Next, we calculate
	\begin{equation}
	\begin{aligned}
	h^{\mathrm{DD}}_{kl,k^{\prime}l^{\prime}} =&  \frac{1}{M} \sum_{i=1}^{P} e^{-j2\pi\nu_{i}\tau_{i}} \Bigg[ \sum_{u=0}^{M-1} \sum_{n=0}^{N-1} \overline{\gamma}_{n}^{i}(u) e^{-j2\pi \frac{n}{N}(k-k^{\prime}-k_{\nu_{i}}-\kappa_{\nu_{i}})} \Bigg] \Bigg[\sum_{m=0}^{M-1} e^{-j2\pi\frac{m}{M}(u-l)} \\ &\cdot \sum_{m^{\prime}=0}^{M-1}e^{j2\pi \frac{m^{\prime}}{M}(u-l^{\prime}-l_{\tau_{i}})}\Bigg].
	\end{aligned}
	\end{equation}
	With the following relations
	\begin{equation}
	\sum_{m=0}^{M-1} e^{-j2\pi\frac{m}{M}(u-l)} = M\delta\left([u-l]_{M}\right),
	\end{equation}
	\begin{equation}
	\sum_{m^{\prime}=0}^{M-1}e^{j2\pi \frac{m^{\prime}}{M}(u-l^{\prime}-l_{\tau_{i}})} = M\delta\left([u-l^{\prime}-l_{\tau_{i}}]_{M}\right),
	\end{equation}
	we get
	\begin{equation}
	\begin{aligned}
	h^{\mathrm{DD}}_{kl,k^{\prime}l^{\prime}} =&  M \sum_{i=1}^{P} e^{-j2\pi\nu_{i}\tau_{i}} \Bigg[\sum_{n=0}^{N-1} \overline{\gamma}_{n}^{i}(l)  e^{-j2\pi \frac{n}{N}(k-k^{\prime}-k_{\nu_{i}}-\kappa_{\nu_{i}})} \Bigg] \delta\left([l-l^{\prime}-l_{\tau_{i}}]_{M}\right).
	\label{eqA3}
	\end{aligned}
	\end{equation}
	Compared with (\ref{eq17}), an extra term $\overline{\gamma}_{n}^{i}(l)$ is occurred. In order to intuitively explain how the time-variant fading acts on the channel response, similar with the case of (\ref{eqA1}), we use the property of DFT
	\begin{equation}
	\begin{aligned}
	h^{\mathrm{DD}}_{kl,k^{\prime}l^{\prime}} =& M \sum_{i=1}^{P} e^{-j2\pi\nu_{i}\tau_{i}} \Bigg[\frac{1}{N}\sum_{n=0}^{N-1} \overline{\gamma}_{n}^{i}(l)  e^{j2\pi \frac{n}{N}(k^{\prime}+k_{\nu_{i}}+\kappa_{\nu_{i}})} e^{-j2\pi k\frac{n}{N}} \Bigg]\delta\left([l-l^{\prime}-l_{\tau_{i}}]_{M}\right)\\ 
	=&  M \sum_{i=1}^{P} e^{-j2\pi\nu_{i}\tau_{i}} \Bigg[\frac{1}{N}\sum_{n=0}^{N-1} \overline{\gamma}_{n}^{i}(l) e^{-j2\pi k\frac{n}{N}} \circledast \sum_{n=0}^{N-1} e^{-j2\pi \frac{n}{N}(k-k^{\prime}-k_{\nu_{i}}-\kappa_{\nu_{i}})} \Bigg]\delta\left([l-l^{\prime}-l_{\tau_{i}}]_{M}\right).
	\end{aligned}
	\end{equation}
	In this way, the received signal $y_{kl}$ can be expressed as
	\begin{equation}
	\begin{aligned}
	y_{kl} =& \frac{1}{N}\sum_{i=1}^{P} e^{-j 2 \pi \nu_{i} \tau_{i}} \sum_{k^{\prime}=0}^{M-1} \Biggl[ \frac{1}{N}\sum_{n=0}^{N-1} \overline{\gamma}_{n}^{i}(l) e^{-j2\pi k\frac{n}{N}}\circledast \beta_{i}(k-k^{\prime})  \Biggr] x\left[k^{\prime},[l-l_{\tau_{i}}]_{M}\right]
	\end{aligned}
	\end{equation}

	\section{Proof of Proposition 2}
	We divide $C_{nm,n^{\prime}m^{\prime}}$ into two parts, $n^{\prime}=n$ and $n^{\prime}=n-1$. Then, the $y_{kl}$ is given by
	\begin{equation}
	\begin{aligned}
	y_{kl}=&\frac{1}{\sqrt{N M}} \sum_{n=0}^{N-1} \sum_{m=0}^{M-1}\Bigg[\sum_{m^{\prime}=0}^{M-1} C_{nm,n m^{\prime}} X_{nm} +\sum_{m^{\prime}=0}^{M-1} C_{nm,(n-1)m^{\prime}} X_{[n-1]_{N}m^{\prime}}\Bigg]e^{-j 2 \pi\left(\frac{n k}{N}-\frac{m l}{M}\right)}
	\end{aligned}
	\end{equation}
	In the following, we calculate the two parts respectively. Using (\ref{eq20}) and (\ref{eq32}), we have that
	\begin{equation}
	\begin{aligned}
	y_{kl}^{\mathrm{ici}} =&\frac{1}{N M} \sum_{k^{\prime}=0}^{N-1} \sum_{l^{\prime}=0}^{M-1} x_{k^{\prime}l^{\prime}}\Bigg[\sum_{n=0}^{N-1} \sum_{m=0}^{M-1} \sum_{m^{\prime}=0}^{M-1} C_{nm,nm^{\prime}} e^{-j 2 \pi n\left(\frac{k-k^{\prime}}{N}\right)} e^{j 2 \pi\left(\frac{m l-m^{\prime} l^{\prime}}{M}\right)}\Bigg] \\
	=& \frac{1}{N M} \sum_{k^{\prime}=0}^{N-1} \sum_{l^{\prime}=0}^{M-1} x_{k^{\prime} l^{\prime}} h_{kl,k^{\prime}l^{\prime}}^{\mathrm{ici}}.
	\end{aligned}
	\end{equation}
	where
	\begin{equation}
	\begin{aligned}
	h_{kl,k^{\prime}l^{\prime}}^{\mathrm{ici}}&=\frac{1}{M} \sum_{n=0}^{N-1} \sum_{m=0}^{M-1} \sum_{m^{\prime}=0}^{M-1}\Bigg[\sum_{i=1}^{P} \sum_{u=0}^{M-1-l_{\tau_{i}}} \overline{\gamma}_{n}^{i}(u+l_{\tau_{i}})   e^{-j 2 \pi\left((m-m^{\prime}) \Delta f-\nu_{i}\right)\left(\frac{u}{M\Delta f}+\tau_{i}\right)} \\
	& \quad \cdot e^{-j 2 \pi\left(\nu_{i}+m^{\prime} \Delta f\right) \tau_{i}}  e^{j 2 \pi \nu_{i} n T}\Bigg]e^{-j 2 \pi n\left(\frac{k-k^{\prime}}{N}\right)} e^{j 2 \pi\left(\frac{m l-m^{\prime} l^{\prime}}{M}\right)} \\
	& =\frac{1}{M}\sum_{i=1}^{P} \Bigg[\!\sum_{u=0}^{M-1-l_{\tau_{i}}} \sum_{n=0}^{N-1} \overline{\gamma}_{n}^{i}(u+l_{\tau_{i}}) e^{-j 2 \pi \frac{n}{N}(k-k^{\prime}-k_{\nu_{i}}-\kappa \nu_{i})} \\ 
	& \quad \cdot e^{j 2 \pi \frac{u}{M}\left(\frac{k_{\nu_{i}}+\kappa_{\nu_{i}}}{N}\right)} \Bigg] \sum_{m=0}^{M-1} e^{-j 2 \pi \frac{m}{M}(u+l_{\tau_{i}}-l)}  \sum_{m^{\prime}=0}^{M-1} e^{j 2 \pi \frac{m}{M}(u-l^{\prime})} \\
	&=M\sum_{i=1}^{P} \Bigg[\!\sum_{u=0}^{M-1-l_{\tau_{i}}} \!\!\sum_{n=0}^{N-1} \overline{\gamma}_{n}^{i}(u+l_{\tau_{i}}) e^{-j 2 \pi \frac{n}{N}(k-k^{\prime}-k_{\nu_{i}}-\kappa \nu_{i})} \\ & \quad \cdot e^{j 2 \pi \frac{u}{M}\left(\frac{k_{\nu_{i}}+\kappa_{\nu_{i}}}{N}\right)} \Bigg] \delta\left([u+l_{\tau_{i}}-l]_{M}\right)\delta\left([u-l^{\prime}]_{M}\right)
	\end{aligned}
	\end{equation}
	Notice that $\delta\left([u+l_{\tau_{i}}-l]_{M}\right)$ is non-zero only when $0 \leq l-l_{\tau_{i}} \leq M-1-l_{\tau_{i}}$. Meanwhile, using (\ref{eqA2}), we have
	\begin{equation}
	\begin{aligned}
	h_{kl,k^{\prime}l^{\prime}}^{\mathrm{ici}}=&M\sum_{i=1}^{P} e^{j 2 \pi \left(\frac{l-l_{\tau_{i}}}{M}\right)\left(\frac{k_{\nu_{i}}+\kappa_{\nu_{i}}}{N}\right)}\Bigg[\frac{1}{N} \sum_{n=0}^{N-1} \overline{\gamma}_{n}^{i}(l) \\ & \quad \cdot e^{-j2\pi k\frac{n}{N}}\circledast \beta_{i}(k-k^{\prime})  \Bigg] \delta\left([l-l^{\prime}-l_{\tau_{i}}]_{M}\right), \quad l \geq l_{\tau_{i}}
	\end{aligned}
	\label{eqA4}
	\end{equation}
	Analogously, the second part $y_{kl}^{\mathrm{isi}}$ is given by
	\begin{equation}
	\begin{aligned}
	y_{kl}^{\mathrm{isi}} =&\frac{1}{N M} \sum_{k^{\prime}=0}^{N-1} \sum_{l^{\prime}=0}^{M-1} x_{k^{\prime}l^{\prime}}\Bigg[\sum_{n=0}^{N-1} \sum_{m=0}^{M-1} \sum_{m^{\prime}=0}^{M-1} C_{nm,(n-1)m^{\prime}} e^{-j 2 \pi \left(\frac{nk-(n-1)k^{\prime}}{N}\right)} e^{j 2 \pi\left(\frac{m l-m^{\prime} l^{\prime}}{M}\right)}\Bigg] \\
	=& \frac{1}{N M} \sum_{k^{\prime}=0}^{N-1} \sum_{l^{\prime}=0}^{M-1} x_{k^{\prime} l^{\prime}} h_{kl,k^{\prime}l^{\prime}}^{\mathrm{ici}}.
	\end{aligned}
	\end{equation}
	where
	\begin{equation}
	\begin{aligned}
	h_{kl,k^{\prime}l^{\prime}}^{\mathrm{isi}}&=\frac{1}{M} \sum_{n=0}^{N-1} \sum_{m=0}^{M-1} \sum_{m^{\prime}=0}^{M-1}\Bigg[\sum_{i=1}^{P} \sum_{u=M-l_{\tau_{i}}}^{M-1} \!\!\!\! \overline{\gamma}_{n}^{i}(u+l_{\tau_{i}}) e^{-j 2 \pi\left((m-m^{\prime}) \Delta f-\nu_{i}\right)\left(\frac{u}{M\Delta f}+\tau_{i} -T\right)}\\  
	& \quad \cdot e^{-j 2 \pi\left(\nu_{i}+m^{\prime} \Delta f\right) (T-\tau_{i})} e^{j 2 \pi \nu_{i} n T}\Bigg]e^{-j 2 \pi n\left(\frac{k-k^{\prime}}{N}\right)} e^{j 2 \pi\left(\frac{m l-m^{\prime} l^{\prime}}{M}\right)} e^{-j2\pi \frac{k^{\prime}}{N}} \\
	& =\frac{1}{M}\sum_{i=1}^{P} \Bigg[\!\sum_{u=M-l_{\tau_{i}}}^{M-1} \sum_{n=0}^{N-1} \overline{\gamma}_{n}^{i}(u+l_{\tau_{i}}) e^{-j 2 \pi \frac{n}{N}(k-k^{\prime}-k_{\nu_{i}}-\kappa \nu_{i})} e^{j 2 \pi \left(\frac{u-M}{M}\right)\left(\frac{k_{\nu_{i}}+\kappa_{\nu_{i}}}{N}\right)} \Bigg]e^{-j2\pi \frac{k^{\prime}}{N}} \\ 
	&\quad \cdot \sum_{m=0}^{M-1} e^{-j 2 \pi \frac{m}{M}(u+l_{\tau_{i}}-l+M)} \sum_{m^{\prime}=0}^{M-1} e^{j 2 \pi \frac{m}{M}(u-l^{\prime})} \\
	&=M\sum_{i=1}^{P} \Bigg[\!\sum_{u=M-l_{\tau_{i}}}^{M-1} \!\!\sum_{n=0}^{N-1} \overline{\gamma}_{n}^{i}(u+l_{\tau_{i}}) e^{-j 2 \pi \frac{n}{N}(k-k^{\prime}-k_{\nu_{i}}-\kappa \nu_{i})} \\ & \quad \cdot e^{j 2 \pi \left(\frac{u-M}{M}\right)\left(\frac{k_{\nu_{i}}+\kappa_{\nu_{i}}}{N}\right)} \Bigg] e^{-j2\pi \frac{k^{\prime}}{N}} \delta\left([u+l_{\tau_{i}}-l]_{M}\right) \delta\left([u-l^{\prime}]_{M}\right)
	\end{aligned}
	\end{equation}
	Here, $\delta\left([u+l_{\tau_{i}}-l]_{M}\right)$ is non-zero only when $0 < l_{\tau_{i}}-l\leq l_{\tau_{i}}$. Similar, we have
	\begin{equation}
	\begin{aligned}
	h_{kl,k^{\prime}l^{\prime}}^{\mathrm{isi}}=&M\sum_{i=1}^{P} e^{-j2\pi \frac{k^{\prime}}{N}}e^{j 2 \pi \left(\frac{l-l_{\tau_{i}}}{M}\right)\left(\frac{k_{\nu_{i}}+\kappa_{\nu_{i}}}{N}\right)}\Bigg[\frac{1}{N} \sum_{n=0}^{N-1} \\ 
	&\quad \overline{\gamma}_{n}^{i}(l) \circledast \beta_{i}(k-k^{\prime})  \Bigg] \delta\left([l-l^{\prime}-l_{\tau_{i}}]_{M}\right), \quad l < l_{\tau_{i}}
	\end{aligned}
	\label{eqA5}
	\end{equation}
	From (\ref{eqA4}) and (\ref{eqA5}), the $y_{kl}$ can be written as
	\begin{equation}
	\begin{aligned}
	y_{kl} = \frac{1}{N}\sum_{i=1}^{p}\sum_{k^{\prime}=0}^{N-1}e^{j 2 \pi \left(\frac{l-l_{\tau_{i}}}{M}\right)\left(\frac{k_{\nu_{i}}+\kappa_{\nu_{i}}}{N}\right)}\alpha_{i}(k,k^{\prime},l)x\left[k^{\prime},[l-l_{\tau_{i}}]_{M}\right]
	\end{aligned}
	\end{equation}
	where
	\begin{equation}
	\begin{aligned}
	\alpha_{i}(k,k^{\prime},l)=
	\begin{dcases}
	\frac{1}{N} \sum_{n=0}^{N-1} \overline{\gamma}_{n}^{i}(l) \circledast \beta_{i}(k-k^{\prime})e^{-j2\pi \frac{k^{\prime}}{N}} \: &l < l_{\tau_{i}}\\ 
	\frac{1}{N} \sum_{n=0}^{N-1} \overline{\gamma}_{n}^{i}(l) \circledast \beta_{i}(k-k^{\prime})  &l \geq l_{\tau_{i}}
	\end{dcases} 
	\end{aligned}
	\end{equation}
	
	\ifCLASSOPTIONcaptionsoff
	\newpage
	\fi
	
	\bibliographystyle{IEEEtran}
	\bibliography{single.bbl}

\begin{thebibliography}{10}
\providecommand{\url}[1]{#1}
\csname url@samestyle\endcsname
\providecommand{\newblock}{\relax}
\providecommand{\bibinfo}[2]{#2}
\providecommand{\BIBentrySTDinterwordspacing}{\spaceskip=0pt\relax}
\providecommand{\BIBentryALTinterwordstretchfactor}{4}
\providecommand{\BIBentryALTinterwordspacing}{\spaceskip=\fontdimen2\font plus
\BIBentryALTinterwordstretchfactor\fontdimen3\font minus
  \fontdimen4\font\relax}
\providecommand{\BIBforeignlanguage}[2]{{%
\expandafter\ifx\csname l@#1\endcsname\relax
\typeout{** WARNING: IEEEtran.bst: No hyphenation pattern has been}%
\typeout{** loaded for the language `#1'. Using the pattern for}%
\typeout{** the default language instead.}%
\else
\language=\csname l@#1\endcsname
\fi
#2}}
\providecommand{\BIBdecl}{\relax}
\BIBdecl

\bibitem{1347350}
{Ke Liu}, T.~{Kadous}, and A.~M. {Sayeed}, ``Orthogonal time-frequency
  signaling over doubly dispersive channels,'' \emph{IEEE Transactions on
  Information Theory}, vol.~50, no.~11, pp. 2583--2603, 2004.

\bibitem{7925924}
R.~{Hadani}, S.~{Rakib}, M.~{Tsatsanis}, A.~{Monk}, A.~J. {Goldsmith}, A.~F.
  {Molisch}, and R.~{Calderbank}, ``Orthogonal time frequency space
  modulation,'' in \emph{2017 IEEE Wireless Communications and Networking
  Conference (WCNC)}, 2017, pp. 1--6.

\bibitem{1802.02623}
R.~Hadani and A.~Monk, ``Otfs: A new generation of modulation addressing the
  challenges of 5g,'' 2018.

\bibitem{8058662}
R.~{Hadani}, S.~{Rakib}, A.~F. {Molisch}, C.~{Ibars}, A.~{Monk},
  M.~{Tsatsanis}, J.~{Delfeld}, A.~{Goldsmith}, and R.~{Calderbank},
  ``Orthogonal time frequency space (otfs) modulation for millimeter-wave
  communications systems,'' in \emph{2017 IEEE MTT-S International Microwave
  Symposium (IMS)}, 2017, pp. 681--683.

\bibitem{2001.02446}
V.~Rangamgari, S.~Tiwari, S.~S. Das, and S.~C. Mondal, ``Otfs: Interleaved ofdm
  with block cp,'' 2020.

\bibitem{8756831}
E.~{Biglieri}, P.~{Raviteja}, and Y.~{Hong}, ``Error performance of orthogonal
  time frequency space (otfs) modulation,'' in \emph{2019 IEEE International
  Conference on Communications Workshops (ICC Workshops)}, 2019, pp. 1--6.

\bibitem{8686339}
G.~D. {Surabhi}, R.~M. {Augustine}, and A.~{Chockalingam}, ``On the diversity
  of uncoded otfs modulation in doubly-dispersive channels,'' \emph{IEEE
  Transactions on Wireless Communications}, vol.~18, no.~6, pp. 3049--3063,
  2019.

\bibitem{8918014}
G.~D. {Surabhi} and A.~{Chockalingam}, ``Low-complexity linear equalization for
  otfs modulation,'' \emph{IEEE Communications Letters}, vol.~24, no.~2, pp.
  330--334, 2020.

\bibitem{8859227}
S.~{Tiwari}, S.~S. {Das}, and V.~{Rangamgari}, ``Low complexity lmmse receiver
  for otfs,'' \emph{IEEE Communications Letters}, vol.~23, no.~12, pp.
  2205--2209, 2019.

\bibitem{9082873}
W.~{Yuan}, Z.~{Wei}, J.~{Yuan}, and D.~W.~K. {Ng}, ``A simple variational bayes
  detector for orthogonal time frequency space (otfs) modulation,'' \emph{IEEE
  Transactions on Vehicular Technology}, vol.~69, no.~7, pp. 7976--7980, 2020.

\bibitem{8424569}
P.~{Raviteja}, K.~T. {Phan}, Y.~{Hong}, and E.~{Viterbo}, ``Interference
  cancellation and iterative detection for orthogonal time frequency space
  modulation,'' \emph{IEEE Transactions on Wireless Communications}, vol.~17,
  no.~10, pp. 6501--6515, 2018.

\bibitem{8761362}
W.~{Shen}, L.~{Dai}, S.~{Han}, I.~{Chih-Lin}, and R.~W. {Heath}, ``Channel
  estimation for orthogonal time frequency space (otfs) massive mimo,'' in
  \emph{ICC 2019 - 2019 IEEE International Conference on Communications (ICC)},
  2019, pp. 1--6.

\bibitem{9109735}
L.~{Gaudio}, M.~{Kobayashi}, G.~{Caire}, and G.~{Colavolpe}, ``On the
  effectiveness of otfs for joint radar parameter estimation and
  communication,'' \emph{IEEE Transactions on Wireless Communications},
  vol.~19, no.~9, pp. 5951--5965, 2020.

\bibitem{995064}
{Byeong-Gwan Iem}, A.~{Papandreou-Suppappola}, and G.~F. {Boudreaux-Bartels},
  ``Wideband weyl symbols for dispersive time-varying processing of systems and
  random signals,'' \emph{IEEE Transactions on Signal Processing}, vol.~50,
  no.~5, pp. 1077--1090, 2002.

\bibitem{730463}
W.~{Kozek} and A.~F. {Molisch}, ``Nonorthogonal pulseshapes for multicarrier
  communications in doubly dispersive channels,'' \emph{IEEE Journal on
  Selected Areas in Communications}, vol.~16, no.~8, pp. 1579--1589, 1998.

\bibitem{5089511}
F.~. {Han} and X.~. {Zhang}, ``Wireless multicarrier digital transmission via
  weyl-heisenberg frames over time-frequency dispersive channels,'' \emph{IEEE
  Transactions on Communications}, vol.~57, no.~6, pp. 1721--1733, 2009.

\bibitem{STROHMER2001243}
T.~Strohmer, ``Approximation of dual gabor frames, window decay, and wireless
  communications,'' \emph{Applied and Computational Harmonic Analysis},
  vol.~11, no.~2, pp. 243--262, 2001.

\bibitem{661517}
V.~{Tarokh}, N.~{Seshadri}, and A.~R. {Calderbank}, ``Space-time codes for high
  data rate wireless communication: performance criterion and code
  construction,'' \emph{IEEE Transactions on Information Theory}, vol.~44,
  no.~2, pp. 744--765, 1998.

\bibitem{MATZ20111}
\BIBentryALTinterwordspacing
G.~Matz and F.~Hlawatsch, ``Chapter 1 - fundamentals of time-varying
  communication channels,'' in \emph{Wireless Communications Over Rapidly
  Time-Varying Channels}, F.~Hlawatsch and G.~Matz, Eds.\hskip 1em plus 0.5em
  minus 0.4em\relax Oxford: Academic Press, 2011, pp. 1--63. [Online].
  Available:
  \url{https://www.sciencedirect.com/science/article/pii/B9780123744838000017}
\BIBentrySTDinterwordspacing

\bibitem{4489232}
P.~{Jung}, ``Pulse shaping, localization and the approximate eigenstructure of
  ltv channels (special paper),'' in \emph{2008 IEEE Wireless Communications
  and Networking Conference}, 2008, pp. 1114--1119.

\bibitem{850678}
I.~E. {Telatar} and D.~N.~C. {Tse}, ``Capacity and mutual information of
  wideband multipath fading channels,'' \emph{IEEE Transactions on Information
  Theory}, vol.~46, no.~4, pp. 1384--1400, 2000.

\bibitem{4103818}
J.~P. {Rybak} and R.~J. {Churchill}, ``Progress in reentry communications,''
  \emph{IEEE Transactions on Aerospace and Electronic Systems}, vol. AES-7,
  no.~5, pp. 879--894, 1971.

\bibitem{frame}
J.~Kovacevic and A.~Chebira, ``An introduction to frames,'' \emph{Foundations
  and Trends in Signal Processing}, vol.~2, pp. 1--94, 02 2008.

\bibitem{850699}
C.~{Lamy} and J.~{Boutros}, ``On random rotations diversity and minimum mse
  decoding of lattices,'' \emph{IEEE Transactions on Information Theory},
  vol.~46, no.~4, pp. 1584--1589, 2000.

\bibitem{140615}
A.~{Wittneben}, ``Basestation modulation diversity for digital simulcast,'' in
  \emph{[1991 Proceedings] 41st IEEE Vehicular Technology Conference}, 1991,
  pp. 848--853.

\bibitem{1317131}
{Weifeng Su}, Z.~{Safar}, and K.~J.~R. {Liu}, ``Diversity analysis of
  space-time modulation over time-correlated rayleigh-fading channels,''
  \emph{IEEE Transactions on Information Theory}, vol.~50, no.~8, pp.
  1832--1840, 2004.

\bibitem{485720}
J.~{Boutros}, E.~{Viterbo}, C.~{Rastello}, and J.~. {Belfiore}, ``Good lattice
  constellations for both rayleigh fading and gaussian channels,'' \emph{IEEE
  Transactions on Information Theory}, vol.~42, no.~2, pp. 502--518, 1996.

\bibitem{681321}
J.~{Boutros} and E.~{Viterbo}, ``Signal space diversity: a power- and
  bandwidth-efficient diversity technique for the rayleigh fading channel,''
  \emph{IEEE Transactions on Information Theory}, vol.~44, no.~4, pp.
  1453--1467, 1998.

\end{thebibliography}
	
\end{document}